\documentclass[journal]{IEEEtran}
\usepackage{amsmath,amsfonts}
\usepackage{algorithmic}
\usepackage{algorithm}
\usepackage{array}
\usepackage[caption=false,font=normalsize,labelfont=sf,textfont=sf]{subfig}
\usepackage{textcomp}
\usepackage{stfloats}
\usepackage{url}
\usepackage{amsmath}
\usepackage{verbatim}
\usepackage{graphicx}
\usepackage{cite}
\usepackage{booktabs}
\usepackage{todonotes}
\usepackage[caption=false]{subfig}
\begin{document}

\title{SA-LSPL:Sequence-Aware Long- and Short- Term Preference Learning for next POI recommendation}

\author{Bin Wang,
        Yan Zhang,
        Yan Ma, 
        Yaohui Jin,
        and~Yanyan~Xu$^*$
\thanks{B. Wang, P. Ding and Yan Ma are with the College of Information, Mechanical, and Electrical Engineering, Shanghai Normal University, Shanghai 201400, China. }
\thanks{Y. Jin and Y. Xu are with MoE Key Laboratory of Artificial Intelligence, AI Institute, and China Institute for Smart Court, Shanghai Jiao Tong University, Shanghai 200240, China. }
\thanks{$\ast${Corresponding author: yanyanxu@sjtu. edu. cn}}
}

\markboth{}%
{Shell \MakeLowercase{\textit{et al.}}: A Sample Article Using IEEEtran.cls for IEEE Journals}


\maketitle

\begin{abstract}
The next Point of Interest (POI) recommendation aims to recommend the next POI for users at a specific time. As users' check-in records can be viewed as a long sequence, methods based on Recurrent Neural Networks (RNNs) have recently shown good applicability to this task. However, existing methods often struggle to fully explore the spatio-temporal correlations and dependencies at the sequence level, and don't take full consideration for various factors influencing users' preferences. To address these issues, we propose a novel approach called Sequence-Aware Long- and Short-Term Preference Learning (SA-LSPL) for next-POI recommendation. We combine various information features to effectively model users' long-term preferences. Specifically, our proposed model uses a multi-modal embedding module to embed diverse check-in details, taking into account both user's personalized preferences and social influences comprehensively. Additionally, we consider explicit spatio-temporal correlations at the sequence level and implicit sequence dependencies. Furthermore, SA-LSPL learns the spatio-temporal correlations of consecutive and non-consecutive visits in the current check-in sequence, as well as transition dependencies between categories, providing a comprehensive capture of user's short-term preferences. Extensive experiments on two real-world datasets demonstrate the superiority of SA-LSPL over state-of-the-art baseline methods.
\end{abstract}

\begin{IEEEkeywords}
Next POI recommendation, Spatio-temporal, Location based services, Attention mechanism
\end{IEEEkeywords}

\section{Introduction}
\IEEEPARstart{W}{ith} the development of location-based social media, people are increasingly willing to record and share their life updates and geographical locations through mobile devices. Therefore, utilizing these geographical location updates (such as check-in records) to understand user preferences for the next actions becomes crucial. As a result, much research has focused on the next POI recommendation problem \cite{zhang2020next,kong2018hst,yu2019adaptive}. Previous studies have developed various models for the next POI recommendation task by leveraging personalized information from different aspects. Early research on next POI recommendation applied Markov chains to model consecutive transitions, such as FPMC \cite{rendle2010factorizing}. Recently, deep learning methods have greatly improved the performance of next POI recommendation. Some works, like STRNN \cite{liu2016predicting} and DCRF \cite{manotumruksa2017deep}, capture users' dynamic short-term preferences \cite{zhao2020go} using RNNs or their various extensions. To further learn the temporal patterns of user preferences, some researchers jointly model users' long-term and short-term preferences, proposing more expressive models \cite{sun2020go,wu2020personalized}. Additionally, due to the great success of the Transformer architecture \cite{vaswani2017attention}, attention mechanisms have also been applied in some works \cite{feng2018deepmove,luo2021stan,guo2020attentional} to better model sequences and achieve good predictive results. In recent years, graph-based methods have utilized graph representation learning \cite{chang2020learning,yang2019revisiting} and graph neural networks (GNNs) \cite{kipf2016semi} to model user preferences \cite{lim2020stp} and spatio-temporal relationships between locations \cite{luo2021stan,dang2022predicting}, obtaining rich representations \cite{rao2022graph,wang2021spatio} to enhance the performance of next location prediction.

These methods have achieved significant success, but there are still challenges hindering further improvement in recommendation performance. The first issue is how to capture the correlation and dependency relationships between sequences. Although RNNs can be used to learn latent representations of check-in records, it is challenging to capture the inter-sequence correlation and dependency over longer time spans. The second issue is that some users have only a limited number of mobility records, making it difficult to learn their travel preferences and behavioral patterns. The third issue is that due to the inherent limitations of RNNs, they can only model continuous activities in user check-in sequences, failing to fully exploit the relationships between non-continuous POI in terms of time, space, and category transitions.

In this paper, we propose sequence-aware long- and short-term preference learning (SA-LSPL) to recommend the next POI for a user, addressing the challenges mentioned above. Our SA-LSPL can train the next POI recommendation model in an end-to-end manner, capturing both the long-term and short-term preferences of users. Within SA-LSPL, to capture users' long-term preferences, we introduce a historical trajectory encoding module based on Bi-LSTM and a non-local network. Additionally, explicit spatio-temporal relationship-based attention mechanisms and self-attention mechanisms are employed to capture relationships between sequences at the sequence level, providing a more comprehensive understanding of users' mobility patterns. Furthermore, to address the issue that some users have only a small number of mobile records, we not only focus on modeling the personalized preferences of users but also construct a social-level check-in behavior similarity matrix. This matrix is utilized to capture the impact of social factors on users' travel preferences, providing a more comprehensive exploration of travel behavior preferences for users with little mobile records. Finally, in order to better explore users' short-term preferences for travel and fully utilize the temporal, spatial, and categorical transition relationships between non-consecutive POIs, we refined the modeling of both consecutive and non-consecutive visits in the current sequence. The behavioral patterns of continuous check-ins are captured by a current sequence encoding module based on LSTM, complemented by an average pooling layer to better retain information from each check-in record. Additionally, we introduce a spatio-temporal-category dilated LSTM, specifically designed to model the spatio-temporal correlations of non-continuous check-ins and the dependency transitions between POI categories in the current sequence.

In summary, the main contributions of this paper are as follows:
\begin{itemize} 
\item To the best of our knowledge, this is the first attempt in the field of next POI recommendation to explicitly model the spatio-temporal relationships at the sequence level based on attention mechanisms, while revealing the implicit correlations and dependencies between sequences.
\item For users' long-term travel preferences, we propose an attention mechanism at the sequence level. This method aims to utilize the various spatio-temporal travel patterns represented by each historical trajectory and capture the correlations and dependencies between trajectories.
\item To model users' short-term travel preferences, we fully account for the spatio-temporal correlation between continuous and non-continuous visits in the current sequence and use a trainable adaptive weight normalisation operation to balance the weights of the two visit modes.
\item Extensive experiments on two real-world datasets demonstrate the superior performance of SA-LSPL. The results indicate that SA-LSPL outperforms existing methods in terms of accuracy.
\end{itemize}

The remaining sections of this paper are organized as follows. We first review related work in Section 2. Then, in Section 3, we describe the definitions and problem statements. Following that, Section 4 introduces our proposed SA-LSPL model. Section 5 presents the experimental results. Finally, in Section 6, we summarize our paper and outline future work. 

\section{Related work}
POI recommendation and prediction are two distinct yet related and extensively studied topics in Location-Based Social Networks (LBSN): the former typically involves learning users' preferences for POIs, while the latter focuses more on recognizing mobility patterns \cite{gao2019predicting}. Models based on Collaborative Filtering (CF), such as Matrix Factorization (MF) \cite{cheng2012fused,li2015rank} and Tensor Factorization (TF) \cite{zheng2010collaborative}, have been widely applied in POI recommendation for learning users' latent preferences. Additionally, early studies employed methods widely used in other consecutive recommendation tasks, such as Markov Chains. FPMC \cite{rendle2010factorizing} and FPMC-LR \cite{cheng2013you} aim to predict the user's next visit based on the factorization of the probability transition matrix. However, Markov Chain-based methods have limitations in capturing long-term dependencies or predicting exploratory human movement. 

Deep learning-based approaches treat the next POI recommendation as a sequence-to-sequence task and achieve better results than traditional methods. ST-RNN \cite{liu2016predicting} is a innovative work that incorporates spatio-temporal features between consecutive human visits into RNN models to predict human mobility. VANext \cite{gao2019predicting} utilizes an RNN to extract potential features of users' short-term mobility behaviors from the current trajectory segment. It also introduces a novel variational attention mechanism to identify periodic features of users' mobility behaviors based on their historical trajectories. However, they fail to adequately consider the impact of users' individualised travel preferences and social factors. DeepMove \cite{feng2018deepmove} employs a multi-module embedding approach to transform sparse features into dense representations and utilizes a historical attention module to retrieve the most relevant historical trajectory information. PLSPL \cite{wu2020personalized} uses standard LSTM models for short-term trajectory mining and general embedding layers to capture users' preferences. But they cannot capture the dynamic personalised preferences of users and it is difficult to consider the spatio-temporal correlation between trajectories. ATST-LSTM \cite{huang2019attention}  inputs spatio-temporal context information at each step into the LSTM network and uses attention mechanisms to selectively use spatio-temporal context information related to historical check-ins. Yet, it fails to fully consider factors such as categories and the impact of non-consecutive check-in behaviors on user travel preferences. LSTPM \cite{sun2020go}, proposed by Ke Sun and others, introduces a non-local network and a geographically expanded LSTM to model users' long-term and short-term preferences. However, it does not study enough the user travel patterns implied by continuous check-in behaviour and does not consider the influence of time as well as category information when exploring non-continuous check-in behaviour. PG$^2$Net \cite{wang2024pg} learns users' group and personalized preferences through spatio-temporal dependencies and attention-based Bi-LSTM but overlooks the influence of non-consecutive check-in behaviors. STAN \cite{luo2021stan}, proposed by Yingtao Luo and others, extracts relative spatio-temporal information between continuous and non-continuous locations using a spatio-temporal attention network. However, it fails to effectively capture the patterns of user travel preferences implied by continuous check-in behaviors. Graph-Flashback \cite{rao2022graph} constructs a powerful user-POI knowledge graph that can be directly used to learn transition patterns between POIs. GCDAN \cite{dang2022predicting}, proposed by Weizhen Dang and others, embeds spatio-temporal points in trajectories into dense representations, considering both consecutive dependencies within a trajectory and correlations between different trajectories. 

Different from the above mentioned methods,our proposed SA-LSPL model, based on Bi-LSTM, integrates users' personalized preferences, social influence, and inter-sequence correlations, capturing both spatio-temporal correlations of continuous and non-continuous visits in the current sequence and category transition dependencies. 

\section{Problem formulation}

We define $U=\left\{u_1,u_2,u_3,...,u_{|U|} \right\}$, $L=\left\{l_1,l_2,l_3,...,l_{|L|} \right\}$ and $C=\left\{c_1,c_2,c_3,...,c_{|C|} \right\}$  as the sets of users, locations, and categories, respectively. Each location $l_i$ is associated with its corresponding latitude and longitude as well as a category. 

Definition 3.1 (Check-in): A check-in record is a tuple $q=(u_i,l_j,c_k,T_m,W_n)$  indicating that user $u_i$ visited a location $l_i$ of category $c_k$ at $T_m$ on day $W_n$. 

Definition 3.2 (Trajectory): A user's entire check-in record is defined as $S=(q_1^u,q_2^u,q_3^u,...) $, where $q_i^u$ is the i-th check-in record of user $u$. We divide the user's entire check-in record into multiple trajectories according to the time interval (e.g., 24 hours). The trajectory containing the prediction target is defined as the current trajectory $S_n$, and the previous trajectories are defined as the historical trajectories ${S_1, S_2, ... , S_{n-1}}$.


Definition 3.3 (Next POI Recommendation): Base on the above definition,the next POI recommendation problem can be defined as follows:given a specific user's sequence of records ${q_1,q_2,q_3,...,q_{t-1}}$ and a set of historical trajectories,the goal is to predict where user $u_i$ is most likely to go  by learning from both the current trajectory and historical trajectories. 

\section{Proposed framework}
\begin{figure*}[ht]
	\centering
		\includegraphics[scale=.7]{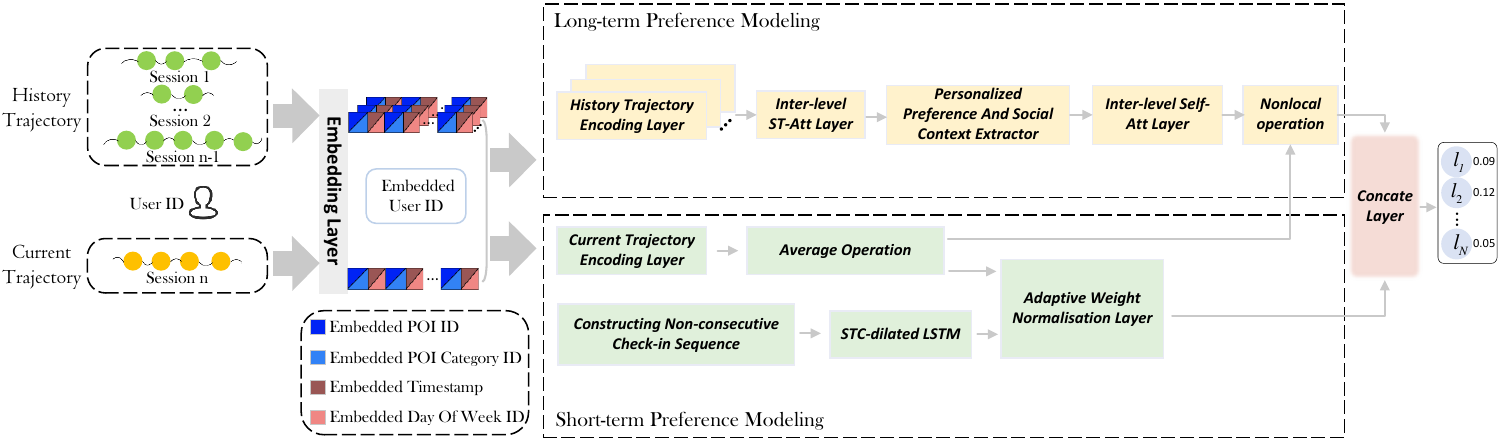}
	\caption{The overall framework of our model.}
	\label{FIG:1}
\end{figure*}

In this section, we provide a detailed overview of our proposed model. Figure \ref{FIG:1} illustrates the architecture of SA-LSPL. It primarily consists of four modules: multi-modal embedding, long and short-term preference modeling, and the prediction component. Our main contribution lies in modeling long and short-term travel preferences. We explore the correlations and dependencies between sequences, along with the influence of social factors. Additionally, we integrate the modeling of non-consecutive check-in behaviors by considering transitions involving temporal, spatial, and categorical factors.

\subsection{\textbf{Multi-modal Embedding Module}}
Trajectory sequences typically contain abundant information about human movement. However, due to mobile device limitations or user behavior, trajectory sequences often exhibit a high degree of sparsity \cite{zhuang2022uncertainty}. To address this issue, we employ sequence embedding methods to handle this kind of data. Taking check-in sequences as an example, they comprise five different types of attributes: user ID, timestamp, day of the week, location, and location category. Different embedding methods are applied to these diverse attributes within the trajectory sequences. 

When dealing with user ID, timestamp, and day of the week, the raw user ID and timestamp cannot be directly input into the model. We refer to the embedding methods mentioned in reference \cite{feng2018deepmove} to handle these two attributes. As each timestamp $t_i$ is continuous and challenging to be embedded directly, we map it to discrete hours. Firstly, we divide a week into 48 time slots, where slots 0-23 represent weekdays and slots 24-47 represent weekends. Each hour is then represented as a one-hot 48-dimensional vector, where non-zero entries indicate the index of the hour. Since one-hot encoding fails to reflect correlations between sequences, we convert it into a dense vector with dimension $D_t$, represented as $V^t \in R^{48 \times D_t}$. For user ID sequences and day of the week sequences, we utilize the same embedding method to map them to dense vectors with dimensions $D_u$ and $D_w$ respectively. The embedding vectors are represented as $V^u \in R^{|U| \times D_u}$ and $V^w \in R^{7 \times D_w}$. 

When handling location and location category, graph embedding (also known as network embedding) has been widely applied in various graph-related research fields in recent years \cite{wang2022graph,kumar2022influence,etaiwi2023semanticgraph2vec}. Given that our research task involves predicting a user's next location, we construct a graph comprising all potential locations an user may reach. To achieve this, we first build a directed weighted graph using the training dataset, where nodes represent locations in the training trajectories, direction follows the appearance order of locations in the trajectories, and weights represent the frequency of consecutive visits between two locations. Subsequently, we employ the graph embedding method node2vec \cite{grover2016node2vec} to map each location to a low-dimensional vector with dimension $D_l$. The embedding vector is represented as $V^l \in R^{|L| \times D_l}$. We adopt the same embedding method for the location category sequence, obtaining an embedding vector represented as $V^c \in R^{|C| \times D_c}$ with dimension $D_c$. This approach allows us to capture features of location interactions. It's worth noting that in the subsequent network training, the embeddings for location and location category will no longer be trained. 

The embedding for each POI containing location, location category, day of the week, and timestamp can be represented as:
 \begin{equation}
    E_i=[V_i^l \oplus V_i^c \oplus V_i^w \oplus V_i^t]
 \end{equation}
where $\oplus$ denotes concatenation, $E_i$ represents the potential vector representation of a POI.
\subsection{\textbf{Long-term Preference Modeling}}
\subsubsection{\textbf{Historical Trajectory Encoding Module}}
Different from the earlier study \cite{sun2020go}, which directly employed LSTM for encoding, we utilize Bi-LSTM to encode historical trajectories. Initially, we embed each POI in each historical trajectory $S_h \in \left\{S_1,S_2,S_3,...,S_{n-1} \right\}$ into a low-dimensional vector. Subsequently, a Bi-LSTM layer is used to learn the high-level representation and temporal dependencies of each POI. 
The calculation process of Bi-LSTM can be described as follows:
 \begin{equation}
   \overrightarrow{h_i} = LSTM(E_i,\overrightarrow{h}_{i-1}),E_i \in S_h
 \end{equation}
 \begin{equation}
   \overleftarrow{h_i} = LSTM(E_i,\overleftarrow{h}_{i-1}),E_i \in S_h
 \end{equation}
 \begin{equation}
   h_i =[\overrightarrow{h}_{i-1} \oplus \overleftarrow{h}_{i-1}]
 \end{equation}
where $h_i$ represents the hidden information of the user's historical trajectory, $\oplus$ denotes concatenation, signifying the combination of the forward and backward outputs. 

\begin{figure*}[ht]   
  \centering            
  \subfloat[\small{Description of spatial correlation}]  
  {
      \label{FIG2:subfig1}\includegraphics[width=0.4\textwidth]{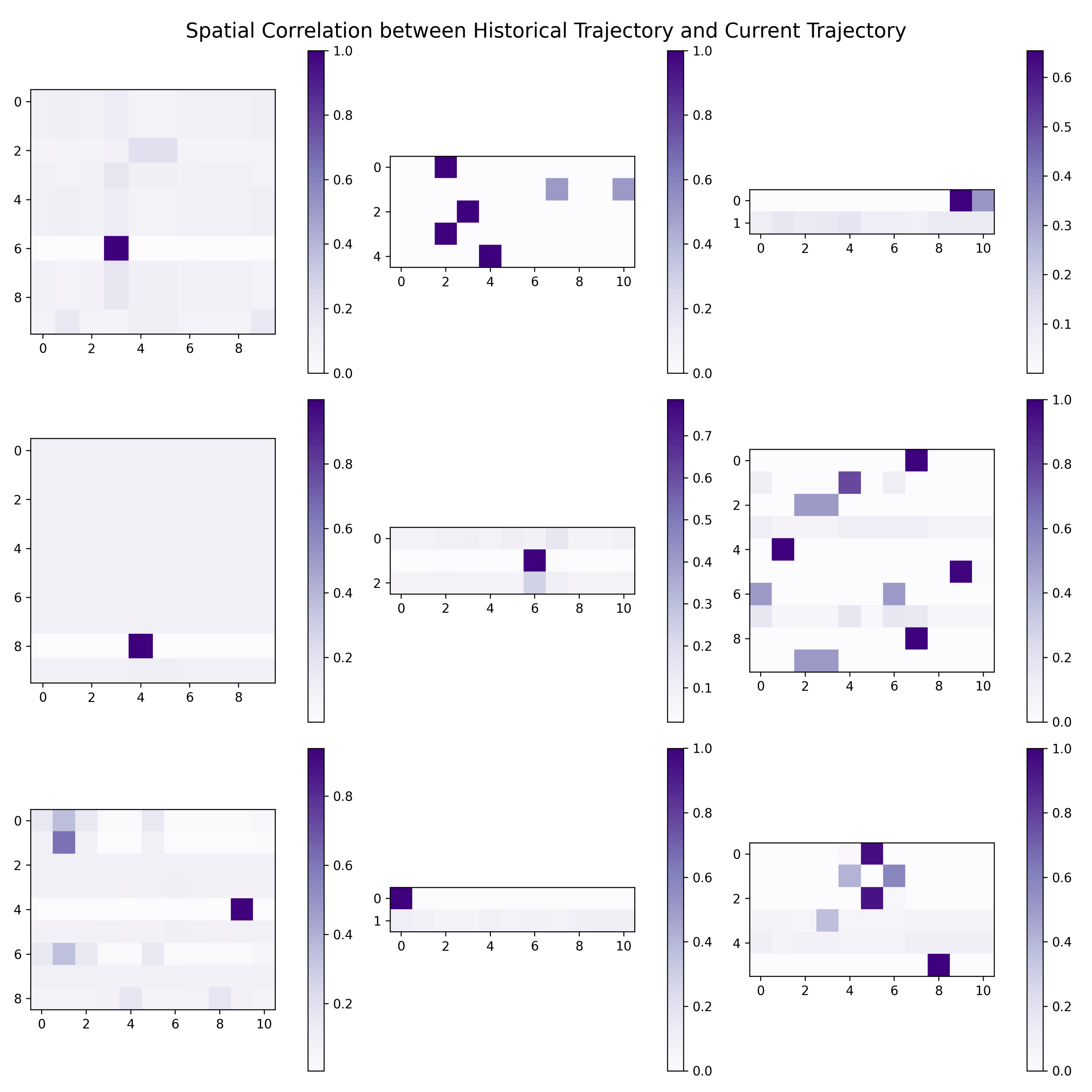}

  }
  \subfloat[\small{Description of temporal correlation}]
  {
      \label{FIG2:subfig2}\includegraphics[width=0.4\textwidth]{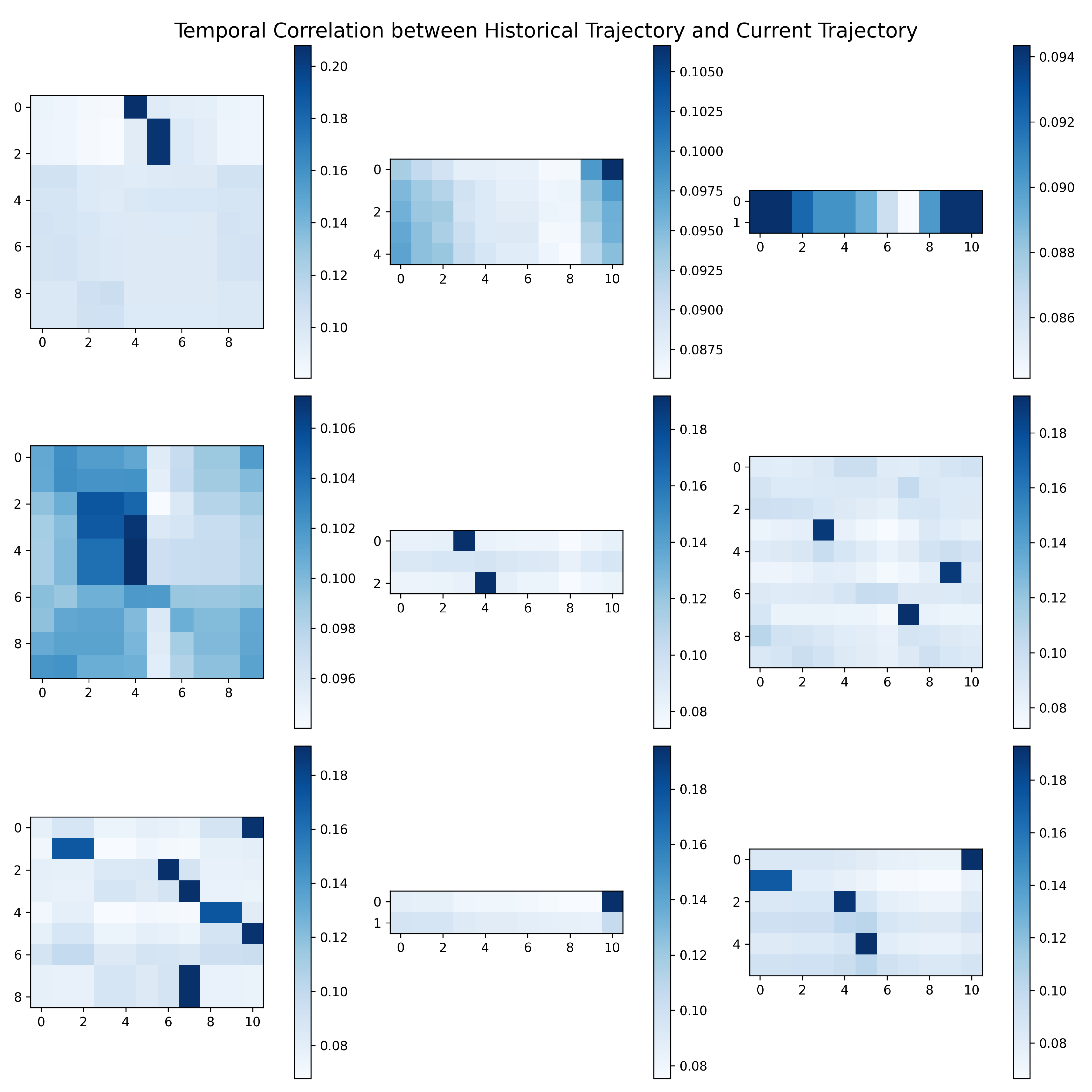}
  }
  \caption{Explanation of spatio-temporal correlation between different trajectories.}   
  \label{FIG:2}            
\end{figure*}

\subsubsection{\textbf{Inter-level ST-Att Layer}}
In general, the distance between geographical locations significantly influences the prediction of a user's next location \cite{sun2020go}. When considering the sequence-level perspective in modeling user distance preferences, the key issue is to select from historical trajectories the one that has the most significant impact on the current trajectory based on the distances between different locations. Meanwhile, the mutual influence between time sequences is also important. For example, users are more inclined to visit restaurants before mealtime in the noon or evening, prefer coffee shops in the afternoon, and might choose bars late at night. We believe that 12:00 noon and 18:00 evening may have a higher correlation, indicating that people may have similar preferences for POIs during these two time periods. This also reflects some group regularities in human travel preferences, which we can utilize to enhance our ability to model user travel preferences.


As shown in Fig. \ref{FIG:2}, we randomly select nine historical trajectories and one current trajectory of a user and explore the spatio-temporal correlation between them. By analysing the relationship between the historical trajectories and the current trajectory, we can reveal the patterns of evolution and spatial changes in user behaviour.
Taking spatial correlation as an example, as shown in Figure \ref{FIG:2}\subref{FIG2:subfig1}, observing the differences in spatial correlation between different locations in different historical trajectories and locations in the current trajectory, we find significant changes. At certain locations, there may be a high degree of spatial correlation between the historical trajectory and the current trajectory, suggesting that users have consistent behavioural patterns at these locations. These locations may represent the user's frequent stops, workplaces, or other important locations. In other locations, the spatial correlation between historical and current trajectories is low, which may imply that the user's behaviour is more variable at these locations.
By investigating the spatio-temporal correlation between different historical and current trajectories, we can better understand users' travel behaviour patterns and spatio-temporal preferences. Based on this study, we propose a method called Inter-level ST-Att Layer\cite{dang2022predicting} to further explore the spatio-temporal correlation between different historical trajectories and current trajectories. The method can capture finer-grained spatio-temporal correlation information between historical trajectories and current trajectories by introducing the Inter-level ST-Att Layer. By combining spatial and temporal features, the Inter-level ST-Att Layer improves the ability to model users' travel behaviour, which helps to mine users' personalised travel preferences and enhance the accuracy of POI recommendation tasks.

Specifically, for spatial correlation, we first generate a geographical distance matrix based on real-world geographical locations and historical trajectories, where the value represents the distance between any two locations. Then, we generate a weight vector between the current trajectory and the historical trajectory based on the distance matrix as follows:
 \begin{equation}
   \alpha_{l_i,l_j} =\frac{exp^{1/dis(l_i,l_j)}}{\sum_{j=1}^{N_h} exp^{1/dis(l_i,l_j)}}
 \end{equation}
 \begin{equation}
   \boldsymbol{\alpha_{h}} = (\alpha_{l_i,l_1},\alpha_{l_i,l_2},...,\alpha_{l_i,l_{N_h}})
 \end{equation}
where $l_i$ is a location in the current trajectory,$l_j$ represents a location in the historical trajectory,$ dis(l_i, l_j)$ is the distance between $l_i$ and $l_j$ , $\alpha_{l_i,l_j}$ measures the impact of location $l_j$ on the current location $l_i$, $N_h$ is the number of locations in the historical trajectory, $\left\{l_1, l_2, ..., l_{N_h}\right\}$ are locations appearing in the historical trajectory, $\boldsymbol{\alpha_{h}}$ is the spatial weight vector between the current location and the historical trajectory. Clearly, the dimension of $\boldsymbol{\alpha_{h}}$ is $1 \times N_h$. 

Meanwhile, to investigate the temporal correlation between different historical trajectories and the current trajectory, similar to the embedding part introduced, we divide a week into 48 time slots, where time slots 0-23 represent weekdays and time slots 24-47 represent weekends. We construct a set of locations to represent the location preferences for each time slot. For example, $Locs_i = \left\{l_1,l_2,l_6,...,l_N \right\}$ is the set of locations appearing in the i-th time slot $T_i $ $i \in (0,47)$. Then, we calculate the time correlation matrix. As shown in Figure \ref{FIG:3}, the time correlation $\tau_{i,j}$ between any two time slots $T_i$ and $T_j$ is expressed as follows: 
\begin{figure}[ht]
	\centering
		\includegraphics[scale=.2]{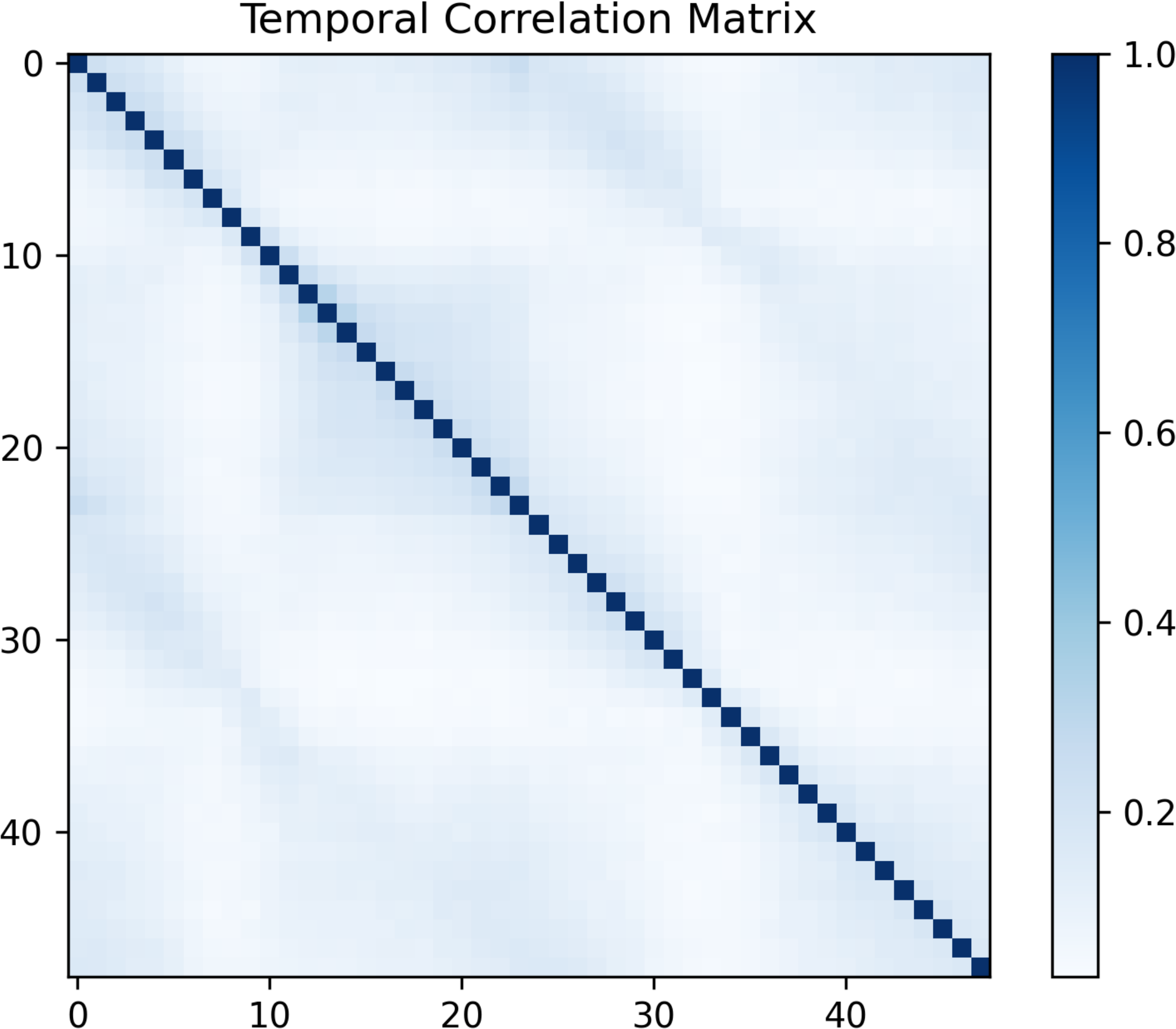}
	\caption{Time correlation matrix.}
	\label{FIG:3}
\end{figure}
 \begin{equation}
   \tau_{i,j} = \frac{|Locs_i \cap Locs_j|}{|Locs_i\cup Locs_j|}
 \end{equation}
Intuitively, the more overlapping POIs two time slots have, the higher their similarity. Finally, we generate a weight vector between the current trajectory and historical trajectories based on the time correlation matrix, as follows:
 \begin{equation}
   \beta_{T_i,T_j} = \frac{exp^{\tau_{i,j}}}{\sum_{j=0}^{47} exp^{\tau_{i,j}}}
 \end{equation}
 \begin{equation}
   \boldsymbol{\beta_{h}} = (\beta_{T_i,T_1},\beta_{T_i,T_2},...,\beta_{T_i,T_{N_h}})
 \end{equation}
where $\tau_{i,j}$ represents the temporal correlation between $T_i$ and $ T_j$, where $T_i$ is the time slot of the current location $l_i$, and $T_j$ is the time slot where the historical location $l_j$ is located. $\beta_{T_i, T_j}$ measures the impact of time slot $T_j$ on time slot $T_i$. $\boldsymbol{\beta_{h}}$ is the time weight vector between the current location and the historical trajectory, generated based on the time-correlation matrix. The dimension of $\boldsymbol{\beta_{h}}$ is $1 \times N_h$. 

Combining the temporal and spatial correlations between different historical trajectories and the current trajectory, we can obtain the user's long-term preference that integrates spatio-temporal context information:
 \begin{equation}
   \mathbf{H_{s,t}} = \boldsymbol{\alpha_{h}} \mathbf{H_{i}} + 
 \boldsymbol{\beta_{h}} \mathbf{H_{i}}
 \end{equation}
where $\mathbf{H_{i}}$ is the output of Bi-LSTM, representing the hidden state of the historical trajectory. The dimension of $\mathbf{H_{i}}$ is $N_h \times K$, where $K$ is the size of the network output vector. 

\subsubsection{\textbf{Personalized Preference And Social Context Extractor}}
When exploring personalized preferences and recognizing potential variations in users' inclinations towards specific POIs, a straightforward approach is to learn each user's location preferences and extract personalized features related to their travel preferences. Simultaneously, human mobility data often grapple with data sparsity issues, wherein certain users have limited movement records \cite{wu2016did}. In such scenarios, social influence emerges as a vital contextual factor to be considered in the subsequent recommendation process for POIs. For example, if the user's friends frequently visit a particular POI location, then that POI may be more attractive to the user. To address this, we employ a direct method to establish a set of friends for identifying user $u_i$. 

\begin{figure}[ht]
	\centering
		\includegraphics[scale=0.24]{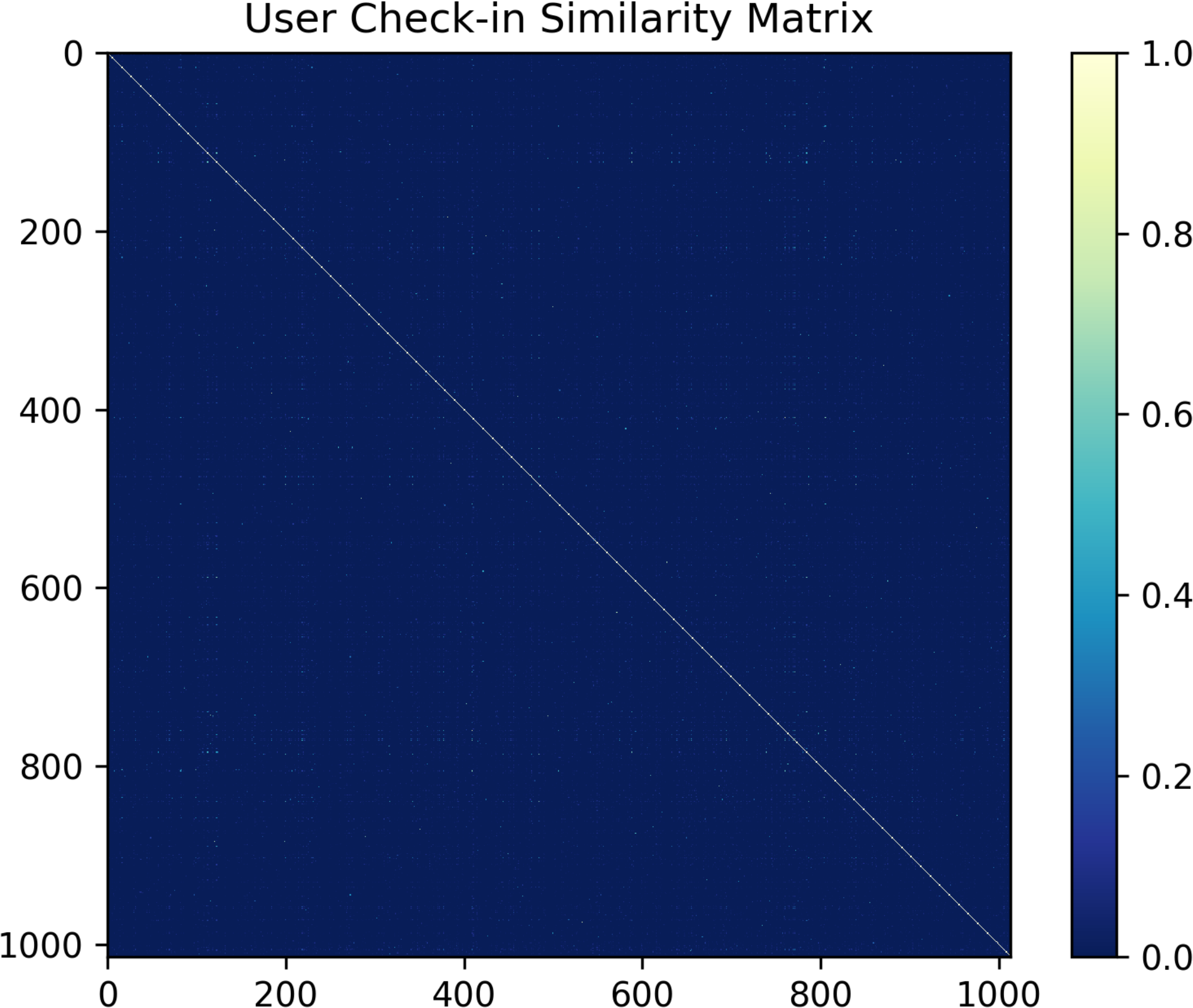}
	\caption{The similarity in check-in behaviors among users.}
	\label{FIG:4}
\end{figure}

The process begins by constructing a check-in vector for each user based on their check-in history. The dimensions of this vector correspond to the number of POIs $|L|$, with each dimension representing the user's visitation frequency to the respective POI. Then we calculate the cosine similarity between the check-in vectors of two users. As shown in Figure \ref{FIG:4}, we have the user check-in behavior similarity matrix, where each element represents the similarity between the check-in behaviors of two users, the effect is more obvious after the picture is enlarged. As each user has their own unique travelling preferences, most users do not show significant similarities, but there are some users who have relatively high similarity of check-ins, which suggests similarity in their travelling preferences. By leveraging this information, we can effectively utilize social context factors to better model user travel preferences. We select the user with the highest similarity to user $u_i$ as the friend of that user. In learning users' personalized location preferences, this helps us better understand the social influences guiding users in the selection of POIs. The process for extracting personalized preferences and social influence is elucidated below:
 \begin{equation}
   \gamma_{u_i} =\frac{exp{\mathbf{({H_i^{\mathrm{T}} V^{u_i}})}}}
   {\sum_{i=1}^{N_h} exp{\mathbf{({H_i^{\mathrm{T}} V^{u_i}})}}}
 \end{equation}
 \begin{equation}
   \gamma_{u_{i,f}} =\frac{exp{\mathbf{({H_i^{\mathrm{T}} V^{u_{i,f}}})}}}
   {\sum_{i=1}^{N_h} exp{\mathbf{({H_i^{\mathrm{T}} V^{u_{i,f}}})}}}
 \end{equation}
 \begin{equation}
  \mathbf{H_{u_i,u_{i_f}}}  =\sum{\gamma_{u_i} \mathbf{H_i}} + \sum{\gamma_{u_{i,f}} \mathbf{H_i}} 
 \end{equation}
where $\mathbf{H_i}$ represents the hidden information of the user's historical trajectory, $\mathbf{V^{u_i}}$ signifies the latent vector of user $u_i$, $\mathbf{V^{u_{i,f}}}$ represents the latent vector of the user whose check-in behavior is most similar to that of user $u_i$, $\gamma_{u_i}$ indicates the importance of each POI for user $u_i$, $\gamma_{u_{i,f}}$ denotes the significance of each POI for the friend of user $u_i$, $N_h$ represents the number of check-in points in a certain historical trajectory, and $\mathbf{H_{u_i,u_{i_f}}}$ is the final representation of the personalized preferences of user $u_i$, incorporating the influence of social factors. 
\subsubsection{Inter-level Self-Att Layer}
To recommend the next POI more effectively, we aim to make full use of the guiding information provided by historical trajectories. Considering that different historical trajectories may reflect users' behavioural changes in different travel, imply different spatio-temporal behavioural patterns, and represent users' travel preferences at different historical stages. So it becomes crucial to utilise the correlation and dependency information between different historical trajectories. To achieve this goal, based on prior research \cite{vaswani2017attention,yang2016hierarchical}, we design a Inter-level Self-attention layer. Its purpose is to comprehensively capture intrinsic correlations and long-range dependency relationships among various historical trajectories while retaining essential information from different historical trajectories. 

In this context, $\mathbf{H_{his}}$ represents a specific historical trajectory with a shape of $(N_h, K)$, $\mathbf{S_{his}}$ represents all historical trajectories with a shape of $(n-1, N_h, K)$, where $n-1$ is the number of historical trajectories, $N_h$ is the number of check-in points in each historical trajectory. We transform the representation of historical trajectories $\mathbf{H_{his}}$ into the representations of query, key, and value through linear transformations, where $\mathbf{W_q}$, $\mathbf{W_k}$, and $\mathbf{W_v}$ are learnable weight matrices. This design aims to achieve a comprehensive understanding of the inherent correlations and long-term dependencies among various historical trajectories. 
 \begin{equation}
   \mathbf{H_{his}= H_{s,t}+H_{u_i,u_{i_f}} }
 \end{equation}
 \begin{equation}
   \mathbf{H_{context}}= Attention(\mathbf{S_{his}W_q,S_{his}W_k,S_{his}W_v})
 \end{equation}
 \begin{equation}
      Attention(Q,K,V)=Softmax(\frac{QK^T}{\sqrt{d_k }})V
 \end{equation}
Leveraging an inter-level self-attention mechanism, this layer captures the correlation, dependency, and key features among different historical trajectories, resulting in a comprehensive representation vector $\mathbf{H_{context}}$ for multiple historical trajectories. 
\subsubsection{\textbf{Nonlocal operation}}
The latent representation of all historical trajectories obtained from the Inter-level Self-Attention Layer is denoted as $\mathbf{H_{context}}$, which includes $\left\{\mathbf{s_1, s_2, ..., s_{n-1} }\right\}$. We employ LSTM to encode the user's current trajectory, followed by an average pooling operation. Considering the correlation between the current situation of the target user and their long-term preferences, we adopt a non-local network that has been proven effective in learning the influence of each historical sequence and the current sequence based on pairwise affinities \cite{sun2020go}. Specifically, we use the following operation to compute the latent representation of the long-term preferences of user $u_i$:
 \begin{equation}
    \mathbf{ h_j} = LSTM(\mathbf{E_i,h_{j-1}}) ,j \in \left\{1,2,...,|S_{n}|\right\}
 \end{equation}
 \begin{equation}
     \mathbf{s_n }=\frac{1}{|S_{n}|} \sum_{j=1}^{|S_{n}|}\mathbf{h_j}
 \end{equation}
 \begin{equation}
     \mathbf{Y_{l}} =\frac{\sum_{i=1}^{n-1} f(\mathbf{s_n,s_i}) \mathbf{W_i s_i} } {\sum_{i=1}^{n-1} f(\mathbf{s_n,s_i})} 
 \end{equation}
 \begin{equation}
    f(\mathbf{s_n,s_i}) = exp(\mathbf{s_n^{\mathrm{T}}s_i})
 \end{equation}
where $\mathbf{s_n}$ is the vector of the current trajectory after an average pooling, preserving all POI information in $S_{n}$, $\mathbf{s_i}$ represents the vector of a historical trajectory. $\mathbf{Y_{l}}$ represents a user's long-term travel preference. The pairwise function $f(\ast)$ calculates an affinity score between the current trajectory $\mathbf{s_n}$ and the historical trajectory $\mathbf{s_i}$, and $\mathbf{W_i}$ is a learnable projection weight matrix. 

\subsection{\textbf{Short-term Preference Modeling}}
\subsubsection{\textbf{Constructing Non-consecutive Check-in Sequence}}

Previous studies \cite{sun2020go,sun2021mfnp} developed Geo-dilated LSTM to capture the geographical influence among non-consecutive check-ins. More recently, a study \cite{chen2022building} extended the consideration to include the temporal cost in addition to capturing the geographical factors, enhancing the overall model performance. However, despite significant progress in these methods, they still do not fully consider the influencing factors between non-consecutive check-ins. Each POI is associated with a category, representing rich semantic information. There exists a important interdependence between categories and POI access behavior, and transitions between different categories are often overlooked. For instance, if a user checks in at an airport POI, the probability of the next visited POI being a home or hotel is much higher than going to a sports stadium or gym. Therefore, considering the category information is essential for the next POI recommendation task.

\begin{figure*}[ht]    

  \centering          
  \subfloat[\small{Category Transition Matrix}]  
  {
      \label{FIG5:subfig1}\includegraphics[scale=0.17]{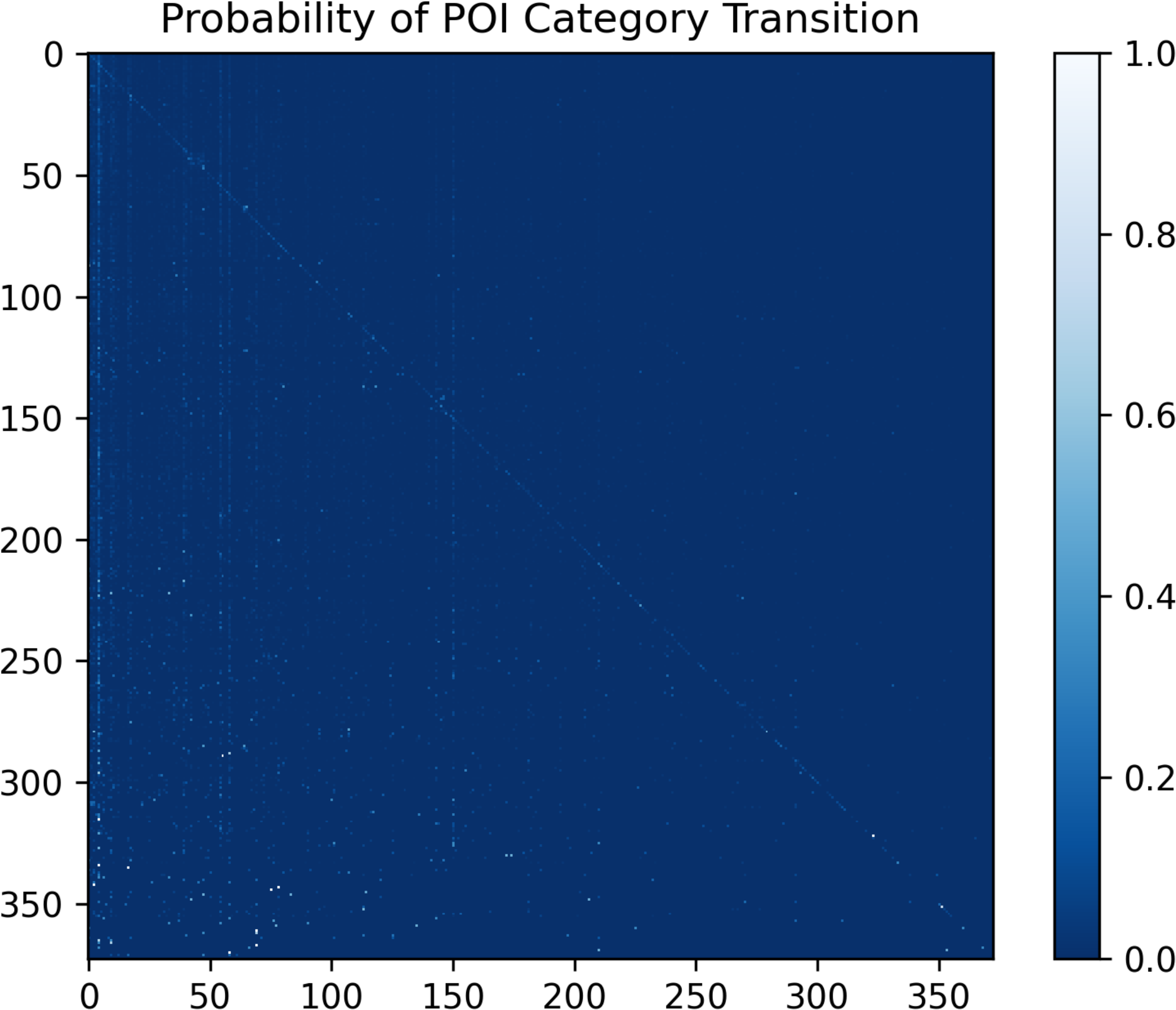} 
  }
  \subfloat[\small{Category Transition Computation Instructions}]
  {
      \label{FIG5:subfig2}\includegraphics[scale=0.86]{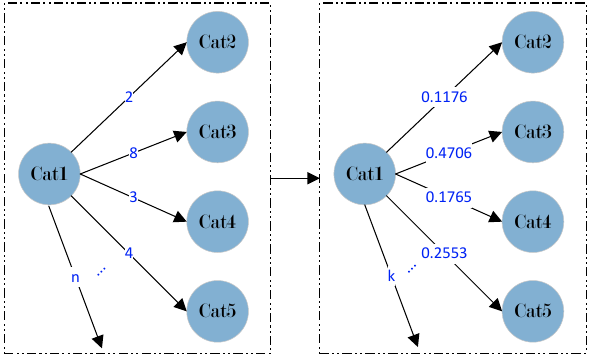}
  }
  \caption{Category Transition Matrix and Generation Process.}    
  \label{FIG:5}         
\end{figure*}

Based on the comprehensive consideration of spatial, temporal and category transition dependencies, we construct a POI category transition relation matrix. The POI category transition relationship matrix based on user trajectories is shown in Figure \ref{FIG:5}\subref{FIG5:subfig1}, the effect is more obvious after the picture is enlarged. For transitions from one category to another, as illustrated in Figure \ref{FIG:5}\subref{FIG5:subfig2}, we conduct a statistical analysis by counting the transition occurrences between different category pairs and normalizing to obtain the transition probabilities between each pair of categories. For example, for Cat1, we first count the number of transitions from Cat1 to each of the other categories. Then, we calculate the softmax scores to obtain the transition probabilities from Cat1 to other categories. Regarding time intervals, similar to constructing the geographical distance matrix, we calculate the average time interval for all transitions from location $l_i$ to $l_j$ if such transitions occurred in any trajectory. Then we normalize both the geographical and time intervals to balance the impact of spatial, temporal, and category transition dependencies. To explore the impact of spatio-temporal-category transition dependencies among non-consecutive POIs in the current trajectory, we select POIs from the check-in sequence as inputs with different skip lengths. The inputs are determined by the spatio-temporal intervals and category transition relationships among POIs in the sequence. For example, given a check-in sequence $S_{n}=\{l1,l2,l3,l4,l5\}$ and a fixed skip length of 2, if $\varepsilon(l1,l3)< \varepsilon(l1,l2)$, then there will be an dilated sequence \{l1,l3\}. Obviously, the smaller the geographical and time intervals, and the greater the category transition probability simultaneously, the more likely it is to generate dilated sequences. We use the following operation to adaptively measure the total cost of geographical distance, time interval, and category transition relationships:

\begin{scriptsize}
    \begin{equation}
    \varepsilon = \! \mathbf{W_1} (\frac{1}{1 + e^{-\Delta dis}}) +  \mathbf{W_2} (\frac{1}{1 + e^{-\Delta tim}}) +  \mathbf{W_3} (1 - \frac{1}{1 + e^{-\Delta cat}})
    \end{equation}
\end{scriptsize}

where $\Delta {dis}$ is the geographical interval based on the normalized spatial matrix, $\Delta {tim}$ is the time interval cost based on the normalized time interval matrix, and $\Delta {cat}$ is the category transition probability based on the category transition matrix. $W_1$, $W_2$, and $W_3$ are trainable weight vectors. Similar operations are applied to check-in POIs l3, l4, and l5, allowing us to construct a new input set. 

\subsubsection{\textbf{STC-dilated LSTM}}

Subsequently, we will collectively consider the set of non-consecutive check-in sequences $S^{STC}_{n}$ generated after taking into account spatial, temporal, and category transformation factors as input for dilated LSTM learning. Based on this, we can obtain the spatio-temporal-category preferences of the user among non-consecutive POIs:
 \begin{equation}
    \mathbf{ h^{'}_{j - 1}}=LSTM(\mathbf{l^{'}_{j - 1}, h^{'}_{j - {\kappa}}})\quad1 < j \leq |S_{n}| 
 \end{equation}
where $\kappa$ is the skip length, automatically determined based on the three influencing factors of spatio-temporal-category transformation. $\mathbf{h^{'}_{j -{\kappa}}}$ is computed from the last sequence $\left\{ \mathbf{l^{'}_\kappa, l^{'}_{j - 1}} \right\} \in S^{STC}_{n}$. 

\subsubsection{\textbf{Adaptive Weight Normalisation Layer}}
At the last check-in timestamp $j - 1$ in the current trajectory, we obtain the latent representations $\mathbf{h_{j - 1}}$ and $\mathbf{h^{'}_{j - 1}}$ learned by the standard and spatio-temporal-category transformation dilated LSTM, respectively. Here, we design an adaptive weight normalization layer, allowing these two vectors to be multiplied by trainable normalization weights individually. This helps the model better adjust the influence of both continuous and non-consecutive sequences on the user's short-term preference. The final representation of short-term user preference is as follows:
 \begin{equation}
    \mathbf{Y_s}=\frac {w_1}{w_1+w_2} \mathbf{h_{j - 1}}  + \frac {w_2}{w_1+w_2} \mathbf{h^{'}_{j - 1}}
\end{equation}
where $w1$ and $w2$ are initialized as 0.5 and adaptively adjusted and updated during the training of the model.And $\mathbf{Y_s}$ combines the preferences for both consecutive and non-consecutive check-in behaviors, preserving the consecutive behavior patterns represented by the user's continuous check-in sequence. It also considers the impact of non-consecutive behavior preferences, incorporating spatial, temporal, and category factors. 

\subsection{\textbf{Prediction}}
By modeling the user's historical trajectories, incorporating spatio-temporal correlations between trajectories, personalized preferences, social influences, and dependencies between trajectories, we obtain the user's long-term preference $\mathbf{Y_l}$. Simultaneously, by merging the influences of both continuous and non-consecutive check-in sequences, we derive the user's short-term preference $\mathbf{Y_s}$. We use the softmax function to compute the probability distribution $p$ for the next POI, as follows:
 \begin{equation}
    y = softmax(\mathbf{W_y (Y_l \oplus Y_s \oplus V^u)}) 
\end{equation}
where $\oplus$ represents the concatenation of personalized preference with long-term and short-term preferences, and $\mathbf{W_y}$ is a trainable parameter matrix. Therefore, the index with the highest probability is used as the next recommended POI, i.e., the place the user is likely to want to visit at the next timestamp. During the model training, we employ negative log likelihood as the loss function. However, in consecutive structure models, the hidden state of the output can more effectively represent the user's latent interests \cite{zhou2019deep}. Therefore, to enhance the predictive accuracy of the network, we propose an auxiliary loss function to supervise the hidden state of the user's target POI \cite{wang2024pg}. The defined loss function is as follows:
 \begin{equation}
    Loss=\frac{1}{L}\left(-\sum_{i=1}^{L} \log(y_i) + \lambda \sum_{i=1}^{L} (v^l_i - \hat{h}_i)^2\right) 
\end{equation}
where $L$ represents the number of samples in the training set, $y_i$ is the output of the softmax layer, $v^l_i$ and $ \hat{h}_i$ are the embeddings of the true next POI and the predicted output of our model, respectively, and \(\lambda\) is a hyperparameter. We choose the L2 loss as the auxiliary loss function. $\lambda$ is used to balance the weights of the prediction and auxiliary loss functions. With the help of the auxiliary loss function, the generated hidden vector can better express the user's interests, leading to improved accuracy in network predictions. 
\section{Experiments}
In this section, we will evaluate the SA-LSPL model on two check-in datasets. We will compare our proposed method with state-of-the-art next POI recommendation models and discuss the experimental results. 
 
 \subsection{\textbf{Experimental Settings}}
 
 We evaluate our model on publicly available Foursquare check-in data collected from New York City (NYC) and Tokyo (TKY) \cite{yang2014modeling}, which is widely used in related studies. The check-in dataset spans approximately 10 months of Foursquare check-in data from NYC and TKY, covering the period from April 12, 2012, to February 16, 2013. The dataset includes anonymized user IDs, location IDs with coordinates, location categories, and timestamps. In our experiments, we exclude POIs in the NYC and TKY datasets that have been visited less than 10 times. Additionally, for each dataset, the check-in records for each user are divided into multiple trajectories based on a 24-hour time window. Each trajectory must contain at least 3 check-ins, and users with fewer than 5 trajectories are filtered out. Finally, we use 80$\%$ of each user's trajectory as the training set, with the remaining 20$\%$ as the test set. The preprocessed dataset statistics are summarized in Table \ref{tbl1}. 

\begin{table*}[ht]
    \centering
    \caption{Description of the datasets.}
    \tabcolsep = 1mm
    \label{tbl1}
    \setlength{\tabcolsep}{1mm}{
    \begin{tabular}{|c|c|c|c|c|c|}
\hline
Dataset& $\#$ of users & $\#$ of POIs & $\#$ of categories & $\#$ of check-ins & $\#$ of trajectories\\ 
\hline
NYC& 1014 & 13994 & 374 & 107071 & 18239 \\ 
\hline
TKY& 2227 & 21052 & 353 & 305050 & 50608  \\
\hline
    \end{tabular}}
\end{table*}

\begin{figure*}[!t]
\centering
\subfloat[\small{The proportional distribution of users' maximum activity radius.}]{\includegraphics[width=3.5in]{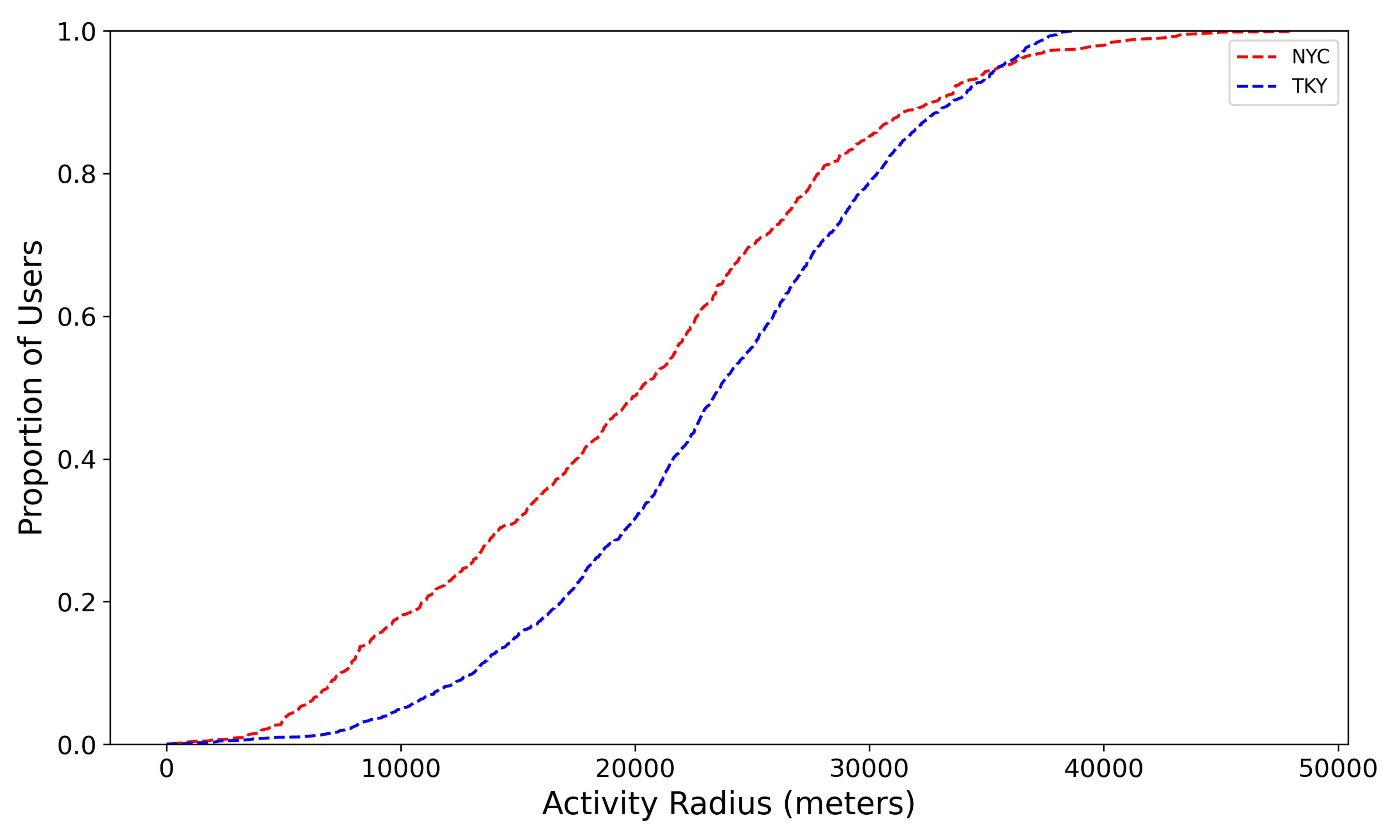}%
\label{FIG6:subfig1}}
\hfil
\subfloat[\small{The proportional distribution of trajectory counts per hour.}]{\includegraphics[width=3.5in]{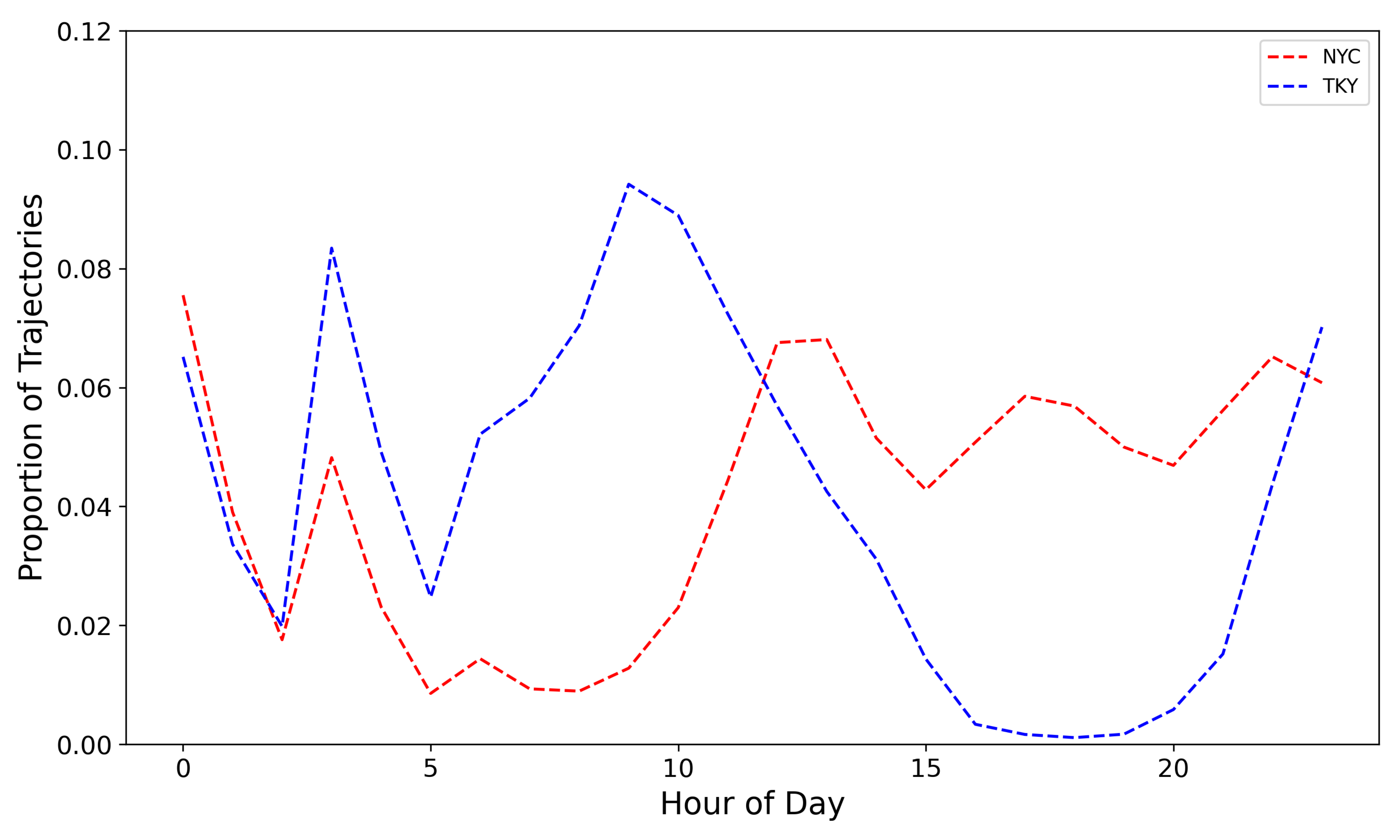}%
\label{FIG6:subfig2}}
\caption{The statistical distributions of the two datasets.}
\label{FIG:6}
\end{figure*}

 Meanwhile, we conduct a brief analysis of the datasets, and more information about the two datasets is shown in Figure \ref{FIG:6}. Figure \ref{FIG:6}\subref{FIG6:subfig1} represents the proportional distribution of users' maximum activity radius, while Figure \ref{FIG:6}\subref{FIG6:subfig2} represents the proportional distribution of trajectory counts per hour. From Figure \ref{FIG:6}\subref{FIG6:subfig1}, we observe that users in NYC and TKY exhibit similar characteristics in terms of their maximum activity radius, with the majority having a range within 30 kilometers. However, due to geographical factors, the maximum activity range for New York users is larger than that of Tokyo users. Figure \ref{FIG:6}\subref{FIG6:subfig2} reveals significant differences in the proportional distribution of trajectory counts per hour between NYC and TKY users, reflecting distinct lifestyle habits in these two locations. 

  

To validate the effectiveness of our proposed method, we compare SA-LSPL with several mainstream deep learning methods:
\begin{itemize} 
\item \textbf{STRNN} \cite{liu2016predicting}:Integrates spatio-temporal context features into the RNN framework to model users' movement behaviors among POIs. 
\item \textbf{LSTM}:A model based on neural networks, it is a variant of RNNs and can efficiently handle consecutive data. 
\item \textbf{Deepmove} \cite{feng2018deepmove}:A neural network model based on attention mechanisms that leverages the historical and current trajectories of each user to learn their preferences.
\item \textbf{STAN} \cite{luo2021stan}:Modeling spatio-temporal correlations between non-adjacent positions through a self-attention network. 
\item \textbf{PLSPL} \cite{wu2020personalized}:A neural network model designed to learn specific preferences for each user, taking into account category information during network construction. 
\item \textbf{PG$^2$Net} \cite{wang2024pg}:A model that learns user group and personalized preferences through spatio-temporal dependencies and attention-based Bi-LSTM. 
\item \textbf{LSTPM} \cite{sun2020go}:This is a state-of-the-art model for next POI recommendation, employing a context-aware non-local network structure and a geographically dilated RNN to capture users' long-term and short-term preferences, respectively. 
\end{itemize}

\begin{figure}[ht]
	\centering
		\includegraphics[scale=.45]{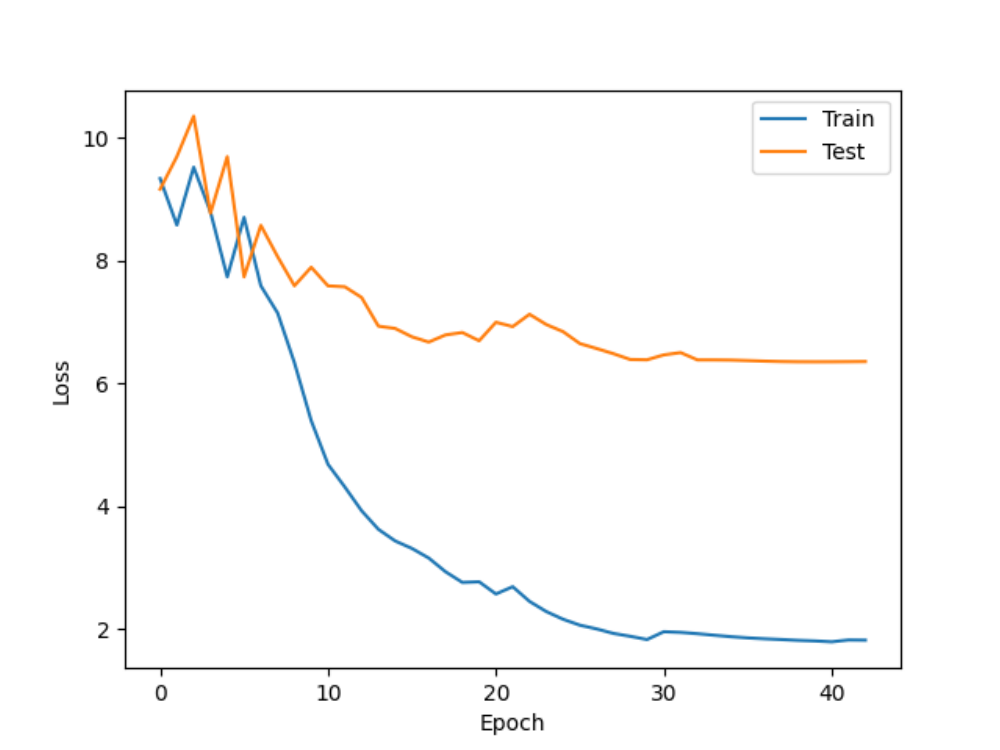}
	\caption{Loss in the training and test process of the NYC dataset.}
	\label{FIG:7}
\end{figure}

For our approach, the embedding dimensions for users and locations are set to $D^u=40$ and $D^l=500$, respectively. We set the embedding dimensions for categories, timestamps, and weekdays as $D^c=50$, $D^t=10$, and $D^w=10$. The hidden state dimension is 500. We use the Adam optimization algorithm for learning all parameters in the model. The initial learning rate and regularization weight are set to 0.0001 and 1e-5, respectively. During training, we employ gradient clipping and adjust the learning rate to ensure optimal model performance. We illustrate the training process of the proposed model using the NYC dataset as an example, and detailed information is available in Figure \ref{FIG:7}. For other baseline models, we set their parameters to the default values provided in the original papers. 

\textbf{Metrics} To compare our model with baseline models, we adopt two evaluation metrics commonly used in previous research \cite{li2018next,huang2020dan}: Recall@K and Normalized Discounted Cumulative Gain (NDCG@K). Recall@K measures whether the correct location is present among the top K recommended POIs. NDCG@K evaluates the quality of the top K recommended POIs. In this paper, we choose K={1,5,10} for comprehensive evaluation. The definitions of Recall@K and NDCG@K are as follows:
 \begin{equation}
  Recall@k = \frac{1}{N} \sum_{u=1}^{N} \frac{|S^k_u \cap S^{visited}_u|}{|S^{visited}_u|} 
 \end{equation}
 \begin{equation}
  NDCG@k= \frac{1}{N} \sum_{u=1}^{N} \frac{1}{z_u} \sum_{j=1}^{k} \frac{2^{I(\left\{S^j_u \right\} \cap S^{visited}_u)} - 1} {\log_2(j+1)} 
 \end{equation}
where $S^k_u$ represents the top $k$ POIs recommended for user $u$, $N$ is the number of users,$S^{visited}_u$ represents the list of visited POIs in the test set, $I(\cdot)$ is an indicator function, $S^j_u$ represents the j-th POI recommended in $S^k_u$, and $z_u$ is the maximum value in DCG@k, which is a normalized constant representing the number of records for which each user makes a prediction. 

\subsection{\textbf{Result and Analysis}}
The experimental results are reported in Table \ref{tbl2}. The results indicate:

\begin{table*}[ht]
    \centering
    \caption{Performance Comparison on NYC and TKY Datasets}
    \label{tbl2}
    \tabcolsep = 1mm
    \begin{tabular}{l|cccccc|cccccc}
        \toprule
        \multicolumn{1}{c|}{\textbf{Model}} &
        \multicolumn{6}{c|}{\textbf{NYC}} &
        \multicolumn{6}{c}{\textbf{TKY}} \\
        \midrule
        & \textbf{Rec@1} & \textbf{Rec@5} & \textbf{Rec@10} & \textbf{NDCG@1} & \textbf{NDCG@5} & \textbf{NDCG@10} &
        \textbf{Rec@1} & \textbf{Rec@5} & \textbf{Rec@10} & \textbf{NDCG@1} & \textbf{NDCG@5} & \textbf{NDCG@10} \\
        \midrule
        STRNN & 0.0936 & 0.1984 & 0.2413 & 0.0936 & 0.1503 & 0.1621 & 0.1232 & 0.262 & 0.316 & 0.1232 & 0.1963 & 0.2137 \\
        LSTM & 0.1014 & 0.2047 & 0.2539 & 0.1014 & 0.1544 & 0.1682 & 0.1357 & 0.271 & 0.3247 & 0.1357 & 0.2072 & 0.2246 \\
        STAN & 0.1203 & 0.3213 & 0.3906 & 0.1203 & 0.2026 & 0.2227 & 0.1311 & 0.2907 & 0.3725 & 0.1311 & 0.1836 & 0.2168 \\
        PLSPL & 0.1241 & 0.3014 & 0.3576 & 0.1241 & 0.2103 & 0.2422 & 0.1278 & 0.3105 & 0.3808 & 0.1278 & 0.223 & 0.2456 \\
        DeepMove & 0.1547 & 0.3209 & 0.3738 & 0.1547 & 0.2425 & 0.2596 & 0.1397 & 0.2745 & 0.3288 & 0.1397 & 0.2107 & 0.2283 \\
        PG$^2$Net & 0.1486 & 0.3251 & 0.3733 & 0.1486 & 0.2427 & 0.2583 & 0.1481 & 0.3094 & 0.3692 & 0.1481 & 0.2336 & 0.2530 \\
        LSTPM & 0.1774 & 0.4035 & 0.4905 & 0.1774 & 0.3001 & 0.3310 & 0.1770 & 0.3955 & 0.4805 & 0.1770 & 0.2915 & 0.3190 \\
        \textbf{SA-LSPL} & \textbf{\pmb{0.1935}} & \textbf{\pmb{0.4086}} & \textbf{\pmb{0.5072}} & \textbf{\pmb{0.1935}} & \textbf{\pmb{0.3067}} & \textbf{\pmb{0.3344}} & \textbf{\pmb{0.2868}} & \textbf{\pmb{0.5123}} & \textbf{\pmb{0.5804}} & \textbf{\pmb{0.2868}} & \textbf{\pmb{0.4082}} & \textbf{\pmb{0.4301}} \\
        \bottomrule
    \end{tabular}
\end{table*}

\begin{itemize} 
\item The proposed SA-LSPL model is compared with baseline models on two datasets, showing overall superior performance. On the NYC and TKY datasets, SA-LSPL outperforms all baseline models across all metrics. Specifically, our method achieves a performance improvement of 9.99\%-26.59\% over STRNN, 9.21\%-25.33\% over LSTM, 3.88\%-13.34\% over DeepMove, 7.32\%-11.66\% over STAN, 6.94\%-14.96\% over PLSPL, 4.49\%-13.39\% over PG$^2$Net, and 1.61\%-1.67\% over LSTPM for Rec@k (k=1,5,10) in the NYC dataset. In terms of NDGC@1, our model improves over STRNN, LSTM, DeepMove, STAN, PLSPL, PG$^2$Net, and LSTPM by 9.99\%, 9.21\%, 3.38\%, 7.32\%, 6.94\%, 4.49\%, and 1.61\%,respectively. On the TKY dataset, our model performs the best across all metrics, with SA-LSPL achieving the highest values for Rec@5, Rec@10, NDCG@5, and NDCG@10. Quantitative evaluation demonstrates the superior effectiveness of our method. 

\item PLSPL outperforms LSTM in all metrics on the NYC and TKY datasets. This is because PLSPL considers contextual information (such as category) to learn specific preferences for each user. However, PLSPL's performance is slightly lower than DeepMove. This phenomenon can be explained by the fact that PLSPL cannot derive useful information from historical trajectories based on the current situation. 
\item Although STAN and LSTPM methods emphasize the importance of non-consecutive check-in POIs in modeling user preferences, their performance is hindered by insufficient exploration of the continuous changes in context (e.g., time and distance) and the lack of consideration for the spatio-temporal context between sequences and the correlation dependence between sequences. This deficiency is a crucial factor in why SA-LSPL outperforms these two methods. Additionally, despite not considering non-consecutive check-in POIs, PLSPL's ability to model personalized weights for users at different locations and DeepMove's consideration of users' periodic behavior make them perform better than STAN in terms of Recall@1, NDCG@1, NDCG@5, and NDCG@10. 
\item Among all baseline methods, the LSTPM model performs best in most metrics, followed by PG$^2$Net. Compared with the PG$^2$Net model, our model also utilizes Bi-LSTM to model long-term user preferences but explicitly considers the spatio-temporal correlation between trajectories at the sequence level. Moreover, we effectively capture implicit correlation and dependence features between trajectories, leading to a better understanding of potential user behavior patterns within sequences. Furthermore, while PG$^2$Net only considers the preferences implied by consecutive check-in actions in modeling short-term user preferences, our model takes into account both consecutive and non-consecutive check-in POIs, balancing the impact of both, which is why our model performs the best on both datasets. 
\end{itemize}

\subsection{\textbf{Ablation Study}}
In this section, we analyze two variants of our proposed SA-LSPL to further evaluate the effectiveness of our model. The two variants are as follows:

\textbf{w/o Short}: A variant model that removes the short-term component SA-LSPL and only utilizes the long-term component.

\textbf{w/o Long}: A variant model that removes the long-term component SA-LSPL and only utilizes the short-term component.

\begin{table*}[ht]
    \centering
    \caption{Ablation Experiment Results}
    \tabcolsep = 1mm
    \label{tbl3}
    \begin{tabular}{l|cccccc|cccccc}
        \toprule
        \multicolumn{1}{c|}{\textbf{Model}} &
        \multicolumn{6}{c|}{\textbf{NYC}} &
        \multicolumn{6}{c}{\textbf{TKY}} \\
        \midrule
        & \textbf{Rec@1} & \textbf{Rec@5} & \textbf{Rec@10} & \textbf{NDCG@1} & \textbf{NDCG@5} & \textbf{NDCG@10} &
        \textbf{Rec@1} & \textbf{Rec@5} & \textbf{Rec@10} & \textbf{NDCG@1} & \textbf{NDCG@5} & \textbf{NDCG@10} \\
        \midrule
        w/o Short & 0.1599 & 0.3867 & 0.4875 & 0.1599 & 0.2784 & 0.3100 & 0.1587 & 0.375 & 0.4508 & 0.1587 & 0.2729 & 0.2975 \\
        w/o Long & 0.1709 & 0.4034 & 0.4972 & 0.1709 & 0.2929 & 0.3261 & 0.1978 & 0.423 & 0.5107 & 0.1978 & 0.316 & 0.3446 \\
        \textbf{SA-LSPL} & \textbf{\pmb{0.1935}} & \textbf{\pmb{0.4086}} & \textbf{\pmb{0.5072}} & \textbf{\pmb{0.1935}} & \textbf{\pmb{0.3067}} & \textbf{\pmb{0.3344}} & \textbf{\pmb{0.2868}} & \textbf{\pmb{0.5123}} & \textbf{\pmb{0.5804}} & \textbf{\pmb{0.2868}} & \textbf{\pmb{0.4082}} & \textbf{\pmb{0.4301}} \\
        \bottomrule
    \end{tabular}
\end{table*}

The experimental results of the ablation study are presented in Table \ref{tbl3}. We observe that our SA-LSPL outperforms both variant models. Specifically:
\begin{itemize} 
\item Overall, the performance of these two variant models on the NYC and TKY datasets is relatively good compared to most baseline models, clearly demonstrating the effectiveness of modeling both long-term and short-term user preferences. 
\item The w/o Short variant model performs poorly on both datasets, especially on the TKY dataset, indicating the importance of capturing user short-term preferences for exploring travel preferences, particularly for users in Tokyo. 
\item The w/o Long variant model performs better than the w/o Short variant model on both datasets and also outperforms most baseline models. This strongly indicates the importance of modeling user short-term preferences for the next POI recommendation. Next, we will further analyze the importance of the key components in modeling user long-term and short-term preferences, respectively. 
\end{itemize}


\subsection{\textbf{Analysis of Key Components in Long-Term Preference Modeling}}

To better understand the impact of each component in modelling users' long-term preferences on network training, we assess the importance of each component using the NYC dataset. In order to compare the impact of social factors, Inter-level ST-Att Layer and Inter-level Self-Att Layer on modelling long-term preferences, we validate the effectiveness of each component by removing them sequentially using a model that does not take short-term preferences into account (w/o Short) as a baseline. Deleting social factors based on the w/o Short model yields the w/o Short\&Socail model, deleting the Inter-level ST-Att Layer yields the w/o Short\&ST-Att model, and deleting the Inter-level Self-Att Layer yields the w/o Short\&Self-Att model. As shown in Table \ref{tbl4}, we evaluate these variant models on three evaluation metrics, NDCG@1, NDCG@5 and NDCG@10, respectively.

\begin{table}[ht]
    \centering
    \caption{Modeling Long-Term Preference Variant
Model Experimental Results.}
    \tabcolsep = 3mm
    \label{tbl4}
    \begin{tabular}{l|ccc}
        \toprule
        
        \textbf{Model} & \textbf{NDCG@1} & \textbf{NDCG@5} & \textbf{NDCG@10} \\
        \midrule
        w/o Short\&Socail & 0.1582 & 0.2758 & 0.3087 \\
        w/o Short\&Self-Att & 0.1553 & 0.2701 & 0.3042 \\
        w/o Short\&ST-Att & 0.1560 & 0.2694 & 0.3021 \\
        w/o Short & \pmb{0.1599} & \pmb{0.2784} & \pmb{0.3100} \\
        \bottomrule
    \end{tabular}
\end{table}

As a whole, the influence of social factors is slightly weaker compared to the Inter-level ST-Att Layer and Inter-level Self-Att Layer modules. The w/o Short\&ST-Att model performs slightly better than the w/o Short\&Self-Att model in terms of NDCG@1, but the w/o Short\&Self-Att model performs better in terms of NDCG@5 and especially NDCG@10. At the same time, we can clearly see that the overall model consisting of the individual components of the w/o Short model works better than the other variants. Overall, the evaluation of these variants of the model demonstrates the validity and respective importance of the individual components in our modelling of users' long-term preferences.

\subsection{\textbf{Analysis of Key Components in Short-Term Preference Modeling}}

Similarly, using the NYC dataset as an example, we evaluate the impact of each component in modelling users' short-term preferences on network training in the same way. We validate the effectiveness of each component by sequentially removing each component based on a model that does not take long-term preferences into account (w/o Long). Deleting the category transition factor based on w/o Long as well as the continuous POI check-in influence, we obtain w/o Long\&C-dilated\&LSTM to study the influence of the category transition factor. Deleting the continuous POI check-in effects yields w/o Long\&LSTM to study the non-continuous POI check-in effects. Deleting the non-continuous POI check-in influence to get w/o Long\&STC-dilated to study the continuous POI check-in influence. The final experimental results of several variants of the model are shown in Table \ref{tbl5}.

From the perspective of each evaluation index, although the influence of category transition factor is relatively weak, it still indicates its effectiveness. At the same time, we can clearly see from each index of NDCG@1, NDCG@5 and NDCG@10 that the influence of continuous POI check-in is much larger than the influence of non-continuous POI check-in. However, through w/o Long we can also see that the combined consideration of both continuous and non-continuous POI check-in makes the model performance more effective, which further illustrates the contribution of STC-dilated LSTM.

\begin{table}[ht]
    \centering
    \caption{ Modeling Short-Term Preference Variant Model Experimental Results.}
    \tabcolsep = 1mm
    \label{tbl5}
    \begin{tabular}{l|ccc}
        \toprule
        \textbf{Model} & \textbf{NDCG@1} & \textbf{NDCG@5} & \textbf{NDCG@10} \\
        \midrule
        w/o Long\&C-dilated\&LSTM & 0.0845 & 0.1436 & 0.1600 \\
        w/o Long\&LSTM & 0.0876 & 0.1471 & 0.1626 \\
        w/o Long\&STC-dilated & 0.1580 & 0.2729 & 0.3074 \\
        w/o Long & \pmb{0.1709} & \pmb{0.2929} & \pmb{0.3261} \\
        \bottomrule
    \end{tabular}
\end{table}

    

\subsection{\textbf{Analysis of Location Embedding Dimension}}

\begin{figure*}[!t]
\centering
\subfloat[\small{Recall@5}]{\includegraphics[width=3.2in]{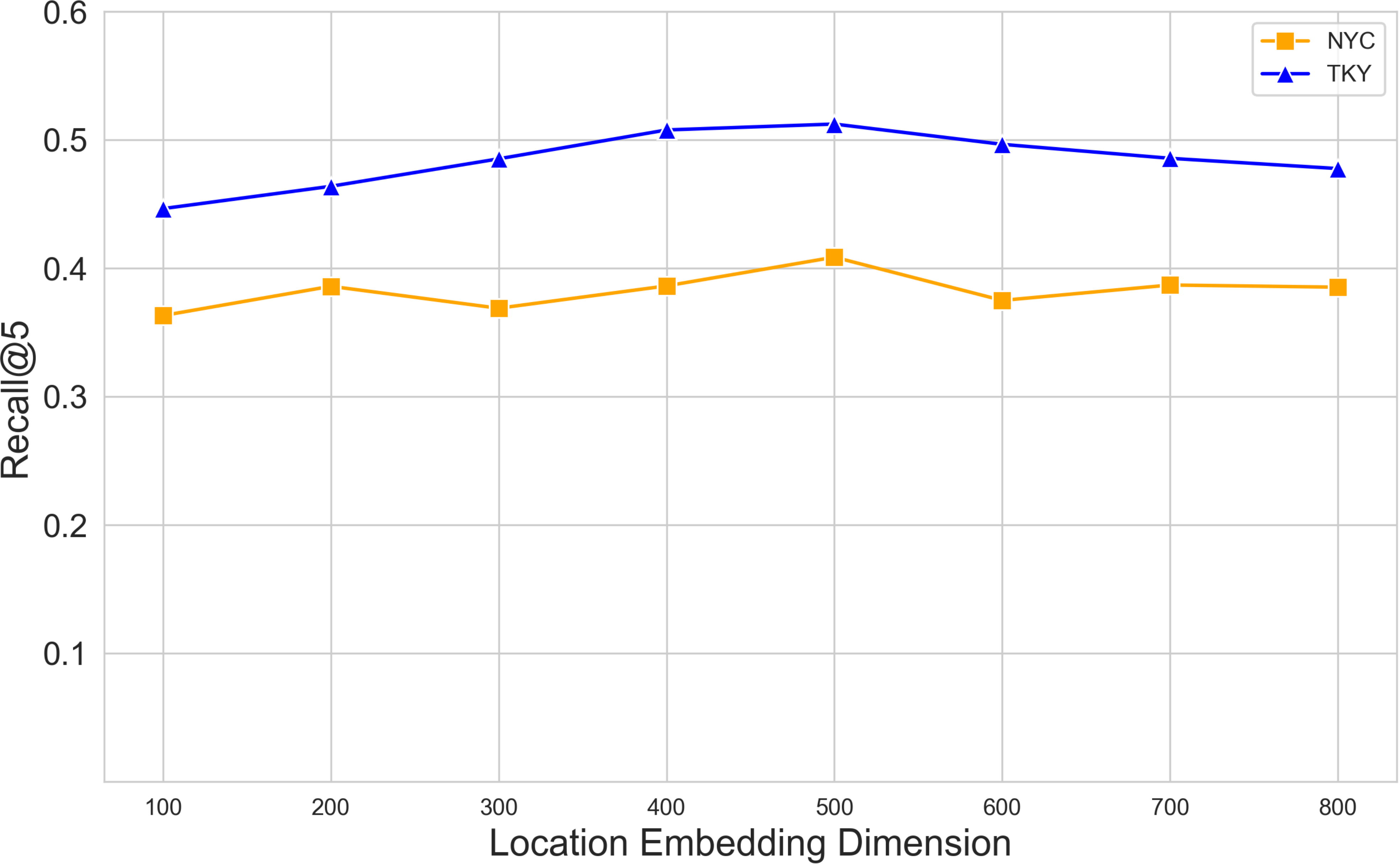}%
\label{FIG8:subfig1}}
\hfil
\subfloat[\small{NDCG@5}]{\includegraphics[width=3.2in]{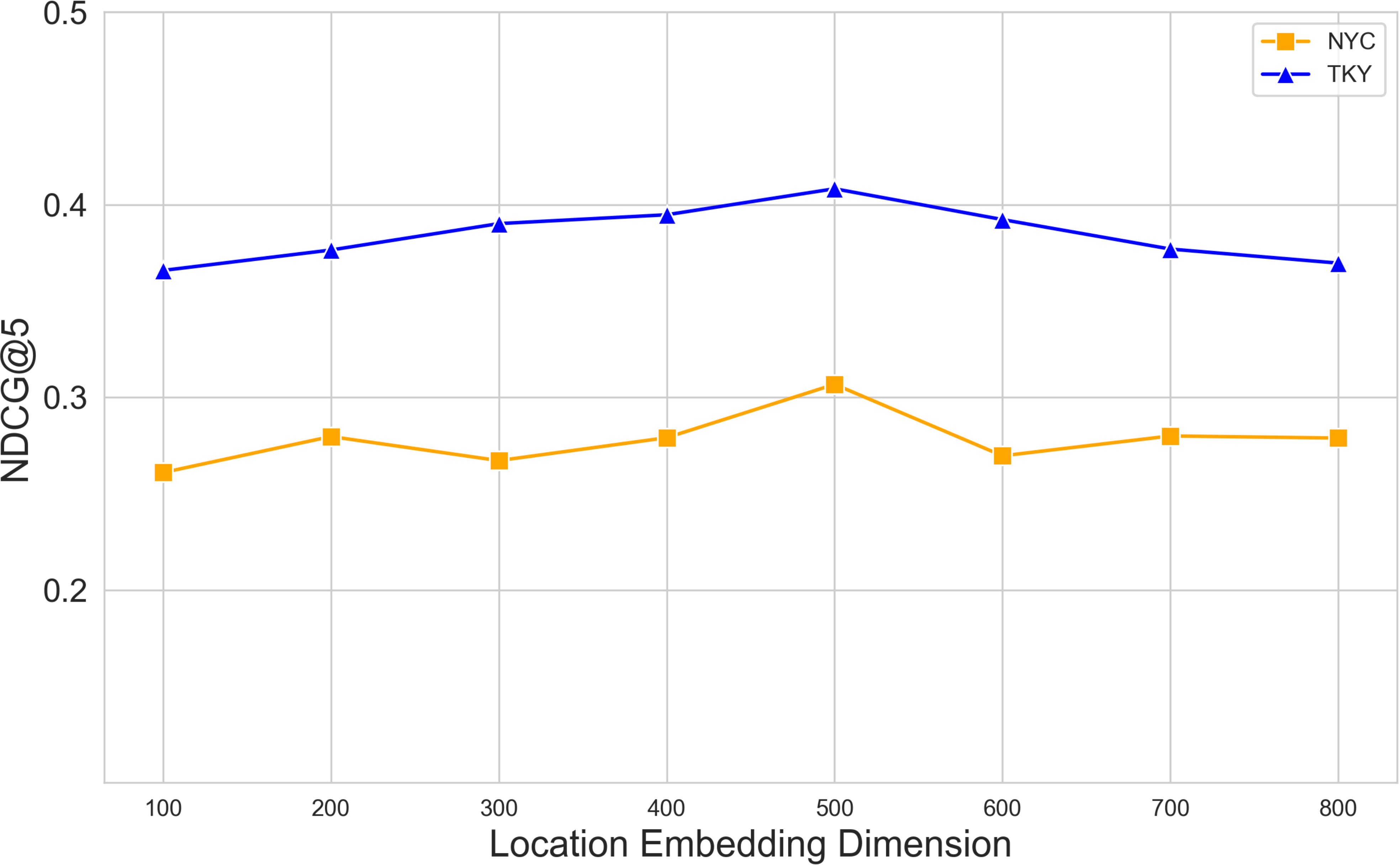}%
\label{FIG8:subfig2}}
\caption{Impact of Location Embedding Dimension.}
\label{FIG:8}
\end{figure*}

To explore the impact of location embedding dimensions, we conduct experiments on two separate datasets. As mentioned earlier, we employ Node2vec to map locations to low-dimensional vectors, which were then fixed and not further involved in training. We vary the dimensions of the embeddings from 100 to 800, with a step size of 100. As shown in Figure \ref{FIG:8}, our model achieved optimal performance when the location embedding dimension was set to $D_l=500$. Additionally, since we perform multimodal embeddings for location, category, user ID, timestamp, and day of the week, the results also indicated that a location embedding dimension of 500 effectively balanced the influence of various features. 

Overall, our model's performance initially increased with the increase in $D_l$ and then slightly declined as d further increased. This is attributed to the fact that smaller location embedding dimensions may lead to an excessive influence of other features, while larger dimensions, although possessing better expressive capabilities, might result in decreased generalization performance and lead to overfitting issues. 

\subsection{\textbf{Case Study}}

\begin{figure*}[!t]
\centering
\subfloat[\small{The check-in trajectory of the historical trajectory most similar to the current trajectory.}]{\includegraphics[width=3.2in]{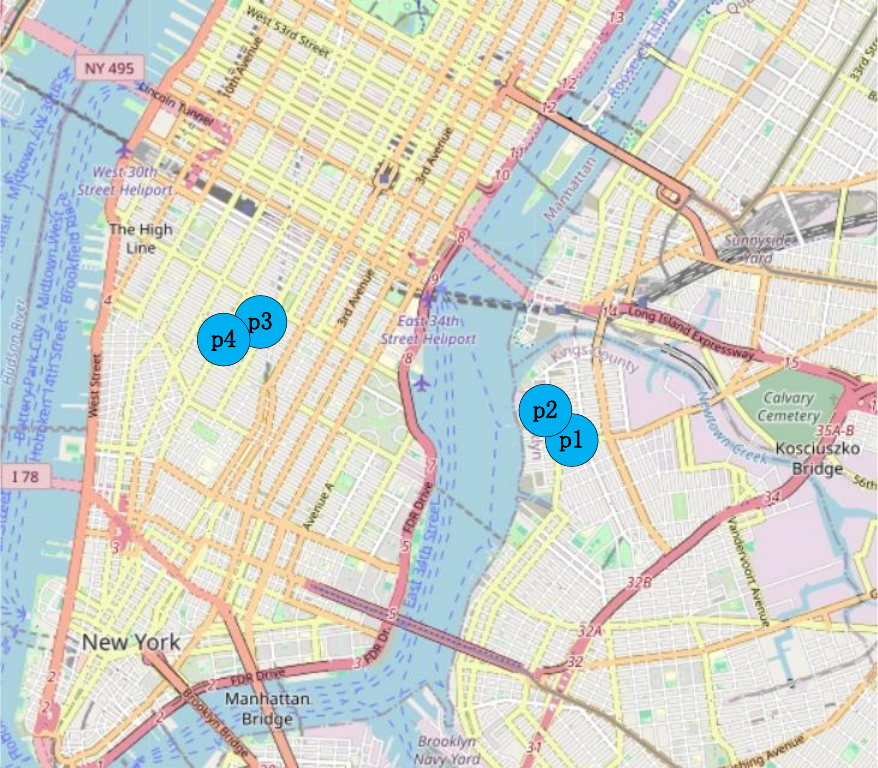}%
\label{FIG9:subfig1}}
\hfil
\subfloat[\small{The check-in trajectory of the current trajectory.}]{\includegraphics[width=3.2in]{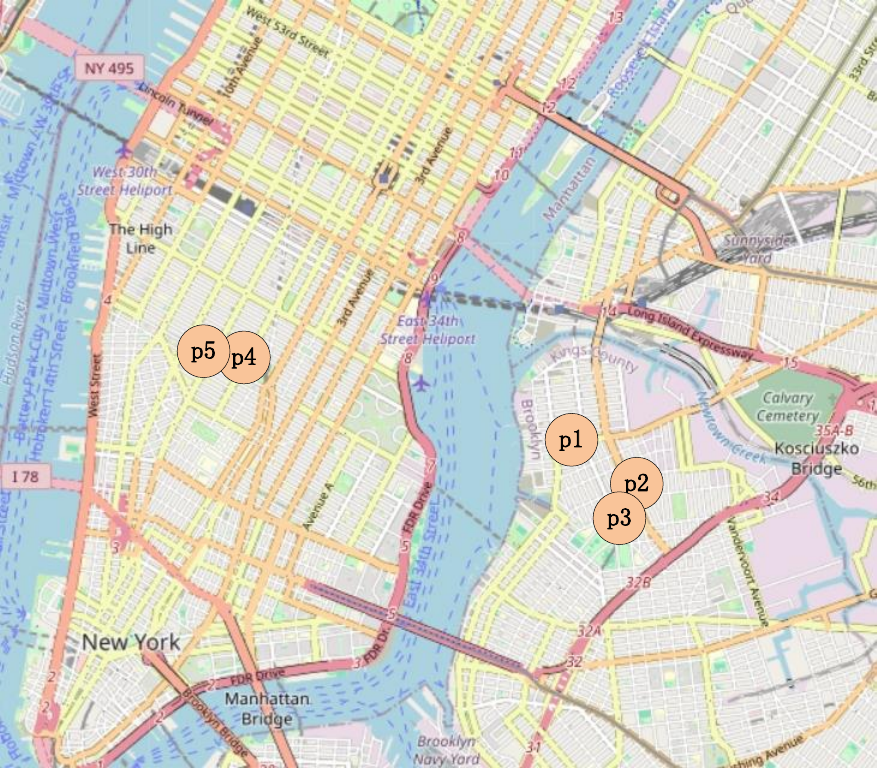}%
\label{FIG9:subfig2}}
\caption{An example explaining how the SA-LSPL model captures user travel preferences and sequence-aware capabilities.}
\label{FIG:9}
\end{figure*}

To enhance the reliability and trustworthiness of our model, emphasizing its sequence-aware capabilities, we conducted a thorough case study by randomly selecting a user from the NYC dataset. Leveraging the pre-trained model, we generated hidden state representations for all historical trajectories and the current trajectory of the chosen user. Through cosine similarity computation, we identify the historical trajectory with the highest similarity to the current trajectory. As shown in Figure \ref{FIG:9}, we refer to the historical trajectory that is most similar to the current trajectory as $S_i$, and the current trajectory as $S_n$. A careful analysis of these two trajectories shows that the model is able to capture the periodicity and behavioural complexity of users' travel preferences very well. 

Specifically, we observe that both the most pertinent historical trajectory $S_i$ and the current trajectory $S_n$ take place on the same day (Tuesday), which is attributed to our incorporation of weekly data into the embedding feature representation. This emphasizes how well our model can understand users' typical recurring behaviors. Furthermore, comparing the check-in paths of $S_i$ and $S_n$ as depicted in Figure \ref{FIG:9}\subref{FIG9:subfig1} and \ref{FIG:9}\subref{FIG9:subfig2}, we see a similarity, suggesting that our model can accurately recognize individual behavior patterns and sequence connections.

\section{\textbf{Conclusion}}
In this paper, we propose a novel end-to-end deep neural network, SA-LSPL, which models users' long-term preferences by comprehensively considering explicit spatio-temporal correlations and implicit dependencies between trajectories. In addition to capturing features from regular consecutive check-in behaviors, we utilize a spatio-temporal-category dilated LSTM to fuse information from non-consecutive check-in POIs, modeling users' short-term preference behaviors. Experimental results demonstrate that our proposed approach significantly improves recommendation accuracy compared to state-of-the-art methods. In future work, we plan to introduce graph neural networks to further explore contextual information for enhancing the performance of next POI recommendation. 

\bibliographystyle{IEEEtranS}
\bibliography{main}

\begin{thebibliography}{10}
\providecommand{\url}[1]{#1}
\csname url@samestyle\endcsname
\providecommand{\newblock}{\relax}
\providecommand{\bibinfo}[2]{#2}
\providecommand{\BIBentrySTDinterwordspacing}{\spaceskip=0pt\relax}
\providecommand{\BIBentryALTinterwordstretchfactor}{4}
\providecommand{\BIBentryALTinterwordspacing}{\spaceskip=\fontdimen2\font plus
\BIBentryALTinterwordstretchfactor\fontdimen3\font minus \fontdimen4\font\relax}
\providecommand{\BIBforeignlanguage}[2]{{%
\expandafter\ifx\csname l@#1\endcsname\relax
\typeout{** WARNING: IEEEtranS.bst: No hyphenation pattern has been}%
\typeout{** loaded for the language `#1'. Using the pattern for}%
\typeout{** the default language instead.}%
\else
\language=\csname l@#1\endcsname
\fi
#2}}
\providecommand{\BIBdecl}{\relax}
\BIBdecl

\bibitem{chang2020learning}
B.~Chang, G.~Jang, S.~Kim, and J.~Kang, ``Learning graph-based geographical latent representation for point-of-interest recommendation,'' in \emph{Proceedings of the 29th ACM International conference on information \& knowledge management}, 2020, pp. 135--144.

\bibitem{chen2022building}
W.~Chen, H.~Wan, S.~Guo, H.~Huang, S.~Zheng, J.~Li, S.~Lin, and Y.~Lin, ``Building and exploiting spatial--temporal knowledge graph for next poi recommendation,'' \emph{Knowledge-Based Systems}, vol. 258, p. 109951, 2022.

\bibitem{cheng2012fused}
C.~Cheng, H.~Yang, I.~King, and M.~Lyu, ``Fused matrix factorization with geographical and social influence in location-based social networks,'' in \emph{Proceedings of the AAAI conference on artificial intelligence}, vol.~26, no.~1, 2012, pp. 17--23.

\bibitem{cheng2013you}
C.~Cheng, H.~Yang, M.~R. Lyu, and I.~King, ``Where you like to go next: Successive point-of-interest recommendation,'' in \emph{Twenty-Third international joint conference on Artificial Intelligence}, 2013.

\bibitem{dang2022predicting}
W.~Dang, H.~Wang, S.~Pan, P.~Zhang, C.~Zhou, X.~Chen, and J.~Wang, ``Predicting human mobility via graph convolutional dual-attentive networks,'' in \emph{Proceedings of the Fifteenth ACM International Conference on Web Search and Data Mining}, 2022, pp. 192--200.

\bibitem{etaiwi2023semanticgraph2vec}
W.~Etaiwi and A.~Awajan, ``Semanticgraph2vec: Semantic graph embedding for text representation,'' \emph{Array}, vol.~17, p. 100276, 2023.

\bibitem{feng2018deepmove}
J.~Feng, Y.~Li, C.~Zhang, F.~Sun, F.~Meng, A.~Guo, and D.~Jin, ``Deepmove: Predicting human mobility with attentional recurrent networks,'' in \emph{Proceedings of the 2018 world wide web conference}, 2018, pp. 1459--1468.

\bibitem{gao2019predicting}
Q.~Gao, F.~Zhou, G.~Trajcevski, K.~Zhang, T.~Zhong, and F.~Zhang, ``Predicting human mobility via variational attention,'' in \emph{The world wide web conference}, 2019, pp. 2750--2756.

\bibitem{grover2016node2vec}
A.~Grover and J.~Leskovec, ``node2vec: Scalable feature learning for networks,'' in \emph{Proceedings of the 22nd ACM SIGKDD international conference on Knowledge discovery and data mining}, 2016, pp. 855--864.

\bibitem{guo2020attentional}
Q.~Guo, Z.~Sun, J.~Zhang, and Y.-L. Theng, ``An attentional recurrent neural network for personalized next location recommendation,'' in \emph{Proceedings of the AAAI Conference on artificial intelligence}, vol.~34, no.~01, 2020, pp. 83--90.

\bibitem{huang2020dan}
L.~Huang, Y.~Ma, Y.~Liu, and K.~He, ``Dan-snr: A deep attentive network for social-aware next point-of-interest recommendation,'' \emph{ACM Transactions on Internet Technology (TOIT)}, vol.~21, no.~1, pp. 1--27, 2020.

\bibitem{huang2019attention}
L.~Huang, Y.~Ma, S.~Wang, and Y.~Liu, ``An attention-based spatiotemporal lstm network for next poi recommendation,'' \emph{IEEE Transactions on Services Computing}, vol.~14, no.~6, pp. 1585--1597, 2019.

\bibitem{kipf2016semi}
T.~N. Kipf and M.~Welling, ``Semi-supervised classification with graph convolutional networks,'' \emph{arXiv preprint arXiv:1609.02907}, 2016.

\bibitem{kong2018hst}
D.~Kong and F.~Wu, ``Hst-lstm: A hierarchical spatial-temporal long-short term memory network for location prediction.'' in \emph{IJCAI}, vol.~18, no.~7, 2018, pp. 2341--2347.

\bibitem{kumar2022influence}
S.~Kumar, A.~Mallik, A.~Khetarpal, and B.~Panda, ``Influence maximization in social networks using graph embedding and graph neural network,'' \emph{Information Sciences}, vol. 607, pp. 1617--1636, 2022.

\bibitem{li2018next}
R.~Li, Y.~Shen, and Y.~Zhu, ``Next point-of-interest recommendation with temporal and multi-level context attention,'' in \emph{2018 IEEE International Conference on Data Mining (ICDM)}.\hskip 1em plus 0.5em minus 0.4em\relax IEEE, 2018, pp. 1110--1115.

\bibitem{li2015rank}
X.~Li, G.~Cong, X.-L. Li, T.-A.~N. Pham, and S.~Krishnaswamy, ``Rank-geofm: A ranking based geographical factorization method for point of interest recommendation,'' in \emph{Proceedings of the 38th international ACM SIGIR conference on research and development in information retrieval}, 2015, pp. 433--442.

\bibitem{lim2020stp}
N.~Lim, B.~Hooi, S.-K. Ng, X.~Wang, Y.~L. Goh, R.~Weng, and J.~Varadarajan, ``Stp-udgat: Spatial-temporal-preference user dimensional graph attention network for next poi recommendation,'' in \emph{Proceedings of the 29th ACM International Conference on Information \& Knowledge Management}, 2020, pp. 845--854.

\bibitem{liu2016predicting}
Q.~Liu, S.~Wu, L.~Wang, and T.~Tan, ``Predicting the next location: A recurrent model with spatial and temporal contexts,'' in \emph{Proceedings of the AAAI conference on artificial intelligence}, vol.~30, no.~1, 2016.

\bibitem{luo2021stan}
Y.~Luo, Q.~Liu, and Z.~Liu, ``Stan: Spatio-temporal attention network for next location recommendation,'' in \emph{Proceedings of the web conference 2021}, 2021, pp. 2177--2185.

\bibitem{manotumruksa2017deep}
J.~Manotumruksa, C.~Macdonald, and I.~Ounis, ``A deep recurrent collaborative filtering framework for venue recommendation,'' in \emph{Proceedings of the 2017 ACM on Conference on Information and Knowledge Management}, 2017, pp. 1429--1438.

\bibitem{rao2022graph}
X.~Rao, L.~Chen, Y.~Liu, S.~Shang, B.~Yao, and P.~Han, ``Graph-flashback network for next location recommendation,'' in \emph{Proceedings of the 28th ACM SIGKDD Conference on Knowledge Discovery and Data Mining}, 2022, pp. 1463--1471.

\bibitem{rendle2010factorizing}
S.~Rendle, C.~Freudenthaler, and L.~Schmidt-Thieme, ``Factorizing personalized markov chains for next-basket recommendation,'' in \emph{Proceedings of the 19th international conference on World wide web}, 2010, pp. 811--820.

\bibitem{sun2021mfnp}
H.~Sun, J.~Xu, K.~Zheng, P.~Zhao, P.~Chao, and X.~Zhou, ``Mfnp: A meta-optimized model for few-shot next poi recommendation.'' in \emph{IJCAI}, 2021, pp. 3017--3023.

\bibitem{sun2020go}
K.~Sun, T.~Qian, T.~Chen, Y.~Liang, Q.~V.~H. Nguyen, and H.~Yin, ``Where to go next: Modeling long-and short-term user preferences for point-of-interest recommendation,'' in \emph{Proceedings of the AAAI Conference on Artificial Intelligence}, vol.~34, no.~01, 2020, pp. 214--221.

\bibitem{vaswani2017attention}
A.~Vaswani, N.~Shazeer, N.~Parmar, J.~Uszkoreit, L.~Jones, A.~N. Gomez, {\L}.~Kaiser, and I.~Polosukhin, ``Attention is all you need,'' \emph{Advances in neural information processing systems}, vol.~30, 2017.

\bibitem{wang2024pg}
B.~Wang, H.~Li, W.~Wang, M.~Wang, Y.~Jin, and Y.~Xu, ``Pg 2 net: Personalized and group preferences guided network for next place prediction,'' \emph{IEEE Transactions on Intelligent Transportation Systems}, 2024.

\bibitem{wang2021spatio}
H.~Wang, Q.~Yu, Y.~Liu, D.~Jin, and Y.~Li, ``Spatio-temporal urban knowledge graph enabled mobility prediction,'' \emph{Proceedings of the ACM on interactive, mobile, wearable and ubiquitous technologies}, vol.~5, no.~4, pp. 1--24, 2021.

\bibitem{wang2022graph}
Z.~Wang, Y.~Zhu, Q.~Zhang, H.~Liu, C.~Wang, and T.~Liu, ``Graph-enhanced spatial-temporal network for next poi recommendation,'' \emph{ACM Transactions on Knowledge Discovery from Data (TKDD)}, vol.~16, no.~6, pp. 1--21, 2022.

\bibitem{wu2016did}
F.~Wu and Z.~Li, ``Where did you go: Personalized annotation of mobility records,'' in \emph{Proceedings of the 25th ACM International on Conference on Information and Knowledge Management}, 2016, pp. 589--598.

\bibitem{wu2020personalized}
Y.~Wu, K.~Li, G.~Zhao, and X.~Qian, ``Personalized long-and short-term preference learning for next poi recommendation,'' \emph{IEEE Transactions on Knowledge and Data Engineering}, vol.~34, no.~4, pp. 1944--1957, 2020.

\bibitem{yang2019revisiting}
D.~Yang, B.~Qu, J.~Yang, and P.~Cudre-Mauroux, ``Revisiting user mobility and social relationships in lbsns: a hypergraph embedding approach,'' in \emph{The world wide web conference}, 2019, pp. 2147--2157.

\bibitem{yang2014modeling}
D.~Yang, D.~Zhang, V.~W. Zheng, and Z.~Yu, ``Modeling user activity preference by leveraging user spatial temporal characteristics in lbsns,'' \emph{IEEE Transactions on Systems, Man, and Cybernetics: Systems}, vol.~45, no.~1, pp. 129--142, 2014.

\bibitem{yang2016hierarchical}
Z.~Yang, D.~Yang, C.~Dyer, X.~He, A.~Smola, and E.~Hovy, ``Hierarchical attention networks for document classification,'' in \emph{Proceedings of the 2016 conference of the North American chapter of the association for computational linguistics: human language technologies}, 2016, pp. 1480--1489.

\bibitem{yu2019adaptive}
Z.~Yu, J.~Lian, A.~Mahmoody, G.~Liu, and X.~Xie, ``Adaptive user modeling with long and short-term preferences for personalized recommendation.'' in \emph{IJCAI}, 2019, pp. 4213--4219.

\bibitem{zhang2020next}
Z.~Zhang, C.~Li, Z.~Wu, A.~Sun, D.~Ye, and X.~Luo, ``Next: a neural network framework for next poi recommendation,'' \emph{Frontiers of Computer Science}, vol.~14, pp. 314--333, 2020.

\bibitem{zhao2020go}
P.~Zhao, A.~Luo, Y.~Liu, J.~Xu, Z.~Li, F.~Zhuang, V.~S. Sheng, and X.~Zhou, ``Where to go next: A spatio-temporal gated network for next poi recommendation,'' \emph{IEEE Transactions on Knowledge and Data Engineering}, vol.~34, no.~5, pp. 2512--2524, 2020.

\bibitem{zheng2010collaborative}
V.~Zheng, B.~Cao, Y.~Zheng, X.~Xie, and Q.~Yang, ``Collaborative filtering meets mobile recommendation: A user-centered approach,'' in \emph{Proceedings of the AAAI Conference on Artificial Intelligence}, vol.~24, no.~1, 2010, pp. 236--241.

\bibitem{zhou2019deep}
G.~Zhou, N.~Mou, Y.~Fan, Q.~Pi, W.~Bian, C.~Zhou, X.~Zhu, and K.~Gai, ``Deep interest evolution network for click-through rate prediction,'' in \emph{Proceedings of the AAAI conference on artificial intelligence}, vol.~33, no.~01, 2019, pp. 5941--5948.

\bibitem{zhuang2022uncertainty}
D.~Zhuang, S.~Wang, H.~Koutsopoulos, and J.~Zhao, ``Uncertainty quantification of sparse travel demand prediction with spatial-temporal graph neural networks,'' in \emph{Proceedings of the 28th ACM SIGKDD Conference on Knowledge Discovery and Data Mining}, 2022, pp. 4639--4647.

\end{thebibliography}
\vfill
\end{document}